\documentstyle[prb,aps,preprint]{revtex}
\begin{document}
\input psfig
\draft

\title{Cotunneling and renormalization effects for the single-electron 
transistor}

\author{J\"urgen K\"onig, Herbert Schoeller and Gerd Sch\"on}

\address{
Institut f\"ur Theoretische Festk\"orperphysik, Universit\"at
Karlsruhe, 76128 Karlsruhe, Germany}

\date{\today}

\maketitle

\begin{abstract}

We study electron transport through a small metallic island in the perturbative 
regime. 
Using a diagrammatic real-time technique, we calculate the occupation of the 
island as well as the conductance through the transistor at arbitrary 
temperature and bias voltage in forth order in the tunneling matrix elements, a 
process referred to as cotunneling. 
Our formulation does not require the introduction of a cutoff. 
At resonance we find significant modifications of previous theories and 
quantitative agreement with recent experiments.
We determine the renormalization of the system parameters and extract
the arguments of the leading logarithmic terms (which can not be derived from
usual renormalization group analysis).
Furthermore, we perform the low- and high-temperature limits.
In the former, we find a behavior characteristic for the multichannel Kondo 
model. 

\end{abstract}
\pacs{73.40.Gk,72.23.Hk,73.40.Rw}

\section{Introduction}

Electron transport through small metallic islands is strongly influenced 
by the charging energy associated with low capacitance of the junctions 
\cite{Ave-Lik,Gra-Dev,Sch-Uebersicht}.
A variety of single-electron effects, including Coulomb blockade phenomena and 
gate-voltage dependent oscillations of the conductance, have been observed. 
If the conductance of  the barriers is low
\begin{equation}
        \alpha_0\equiv h/(4\pi^2 e^2 R_{\rm T}) \ll 1
\end{equation}
they can be described within the ``orthodox theory''\cite{Ave-Lik} which 
treats tunneling in lowest order perturbation theory (golden rule). This
corresponds to the classical picture of incoherent, sequential 
tunneling processes. 
On the other hand, there is
experimental and theoretical evidence that in several regimes higher-order 
tunneling processes have to be taken into account. 

First, in the Coulomb blockade regime, sequential tunneling is exponentially
suppressed. 
The leading contribution to the current is a second-order process in $\alpha_0$ 
where electrons tunnel via a virtual state of the island.
In Ref.~\onlinecite{Ave-Naz} the transition rate of this 
``inelastic cotunneling'' process was evaluated at zero temperature.
A divergence arises at finite temperature which requires a regularization.
In Ref.~\onlinecite{Ave-Naz} this was done within an approximation which is 
valid far away from the resonances. 
In this regime, their results have been confirmed by experiments \cite{Exp-Cot}.

Second, at resonance, even though sequential tunneling occurs, higher-order 
processes have a significant effect on the gate-voltage dependent linear and 
nonlinear conductance \cite{SS,KSS1,KSS3}. 
The energy gap between two adjacent charge states as well as the tunneling 
conductance is renormalized.
Similar effects have been discussed for the equilibrium properties of the 
single-electron box \cite{Laf,Mat,Gol,Fal-Schoen-Zim,Gra}.  
A diagrammatic real-time technique developed for metallic islands 
\cite{SS,KSS1} as well as for quantum dots \cite{KSS2,KSSS} allows a systematic 
description of the nonequilibrium tunneling processes. 
The effects of quantum fluctuations become observable either for strong 
tunneling $\alpha_0\sim 1$ or at low temperatures $\alpha_0 \ln ({E_{\rm C}/T}) 
\sim 1$, where $E_{\rm C}$ denotes the charging energy (see below). 
The theory has been evaluated in Ref.~\onlinecite{SS,KSS1} in the limit where 
only two adjacent charge states are included (even virtually).
Therefore, it was necessary to introduce a bandwidth cutoff $\sim E_{\rm C}$. 
The predicted broadening of the conductance peak as well as the reduction of its
height has been confirmed qualitatively in recent experiments on a 
single-electron transistor in the strong tunneling regime by 
Joyez {\sl et al.} \cite{esteve}. 
However, a quantitative fit between theory and experiment requires introducing a 
renormalized value for the charging energy. 

In Ref.~\onlinecite{KSS3}, we have used the real-time diagrammatic technique to 
obtain the current in second order in $\alpha_0$ at low temperature when only 
two adjacent charge states are classically occupied, but {\it all} relevant 
virtual states are included.
In this case no cutoff is required; all terms are regularized in a natural way.
This analysis allows an unambiguous comparison with experiments where only bare 
system parameters enter.
At resonance we obtain new contributions compared to the earlier theory of 
electron cotunneling. 
They emerge from a renormalization of the charge excitation energy and the 
tunneling conductance.
For realistic parameters $T/E_{\rm C}\sim 0.05$ and 
$\alpha_0^{\rm L}=\alpha_0^{\rm R} \sim 0.02$ the corrections are of order 
$20\%$. 

In this article, we generalize the analysis of Ref.~\onlinecite{KSS3} in the 
following points:\\ 
-- We allow for arbitrary temperature.
Our theory is, therefore, applicable both in the Coulomb blockade regime at 
low temperature as well as the classical regime at high temperature.
While the effects of cotunneling are most dramatic in the first case, signatures
of the higher order contributions are visible even in the second case.\\ 
-- We allow for arbitrary transport voltage, i.e., also extreme non-equilibrium
situations are included in which more than two charge states are classically 
occupied.\\
-- We present analytic expressions for the current, and we discuss the asymptotic
behavior at low and high temperatures.
Renormalization group studies of the single-electron box 
\cite{Mat,Fal-Schoen-Zim} predict a renormalization of the system parameters, 
which depends logarithmically on the relevant energy scale.
This renormalization shows up in our formulas in new terms which are crucial
at resonance.
We are, therefore, able to extract the argument of the leading logarithmic terms.
Finally, we compare with recent experiments \cite{esteve} and find very good 
agreement without any fitting parameter.

This paper is organized as follows:

In Sec.~\ref{section model} we define the model, which is based on the standard 
tunneling Hamiltonian, and describe the diagrammatic technique which we use
in the following.
We, then, set up the general scheme for a systematic perturbation expansion in 
the dimensionless conductance $\alpha_0$ in Sec.~\ref{section expansion}.

Sec.~\ref{section general} contains the general result for the expansion up
to second order, in which we find four terms contributing to the total current
$I^{(2)} = I^{(2)}_1+I^{(2)}_2+I^{(2)}_3+I^{(2)}_4$, 
Eqs.~(\ref{general 1})-(\ref{general 4}).
The first one, $I^{(2)}_1$, is related to processes in which one electron enters 
and another one leaves the island coherently.
This term is responsible for a finite transport in the Coulomb blockade regime.
We recover the usual ''cotunneling'' \cite{Ave-Naz} but without any 
regularization problems at finite temperature.
The second and third contribution, $I^{(2)}_2$ and $I^{(2)}_3$, describe
the renormalization of the dimensionless conductance and of the energy gap for 
adjacent charge states, respectively.
These terms are important at resonance.
The fourth contribution, $I^{(2)}_4$, corresponds to processes in which two 
electrons enter or leave the island coherently.
This plays a role in situations in which more than two charge states are 
classically occupied, i.e., at high temperature or high voltage.
For a quantitative comparison with experiments \cite{esteve} {\it all} terms 
are important.

The results simplify to closed analytic expressions in the two most interesting 
cases.
The first is the low temperature regime (Sec.~\ref{section low temp}) in which
the renormalization of the system parameters plays an important role.
The dimensionless conductance as well as the energy gap between adjacent charge
states are effectively reduced.
A discussion of this renormalization and a comparison is given in 
Sec.~\ref{section rg}.

The other case in which we obtain closed analytic expressions is the linear 
response regime for arbitrary temperature (Sec.~\ref{section all temp}).
We finally perform an expansion for low and high temperatures, respectively,
and obtain very simple results for these limits in Sec.~\ref{section asymptotic}.

In general a numerical solution of a set of equations is necessary to evaluate 
the contribution $I^{(2)}_2$.

\section{The Model and diagrammatic technique}\label{section model}

\begin{figure}
\centerline{\psfig{figure=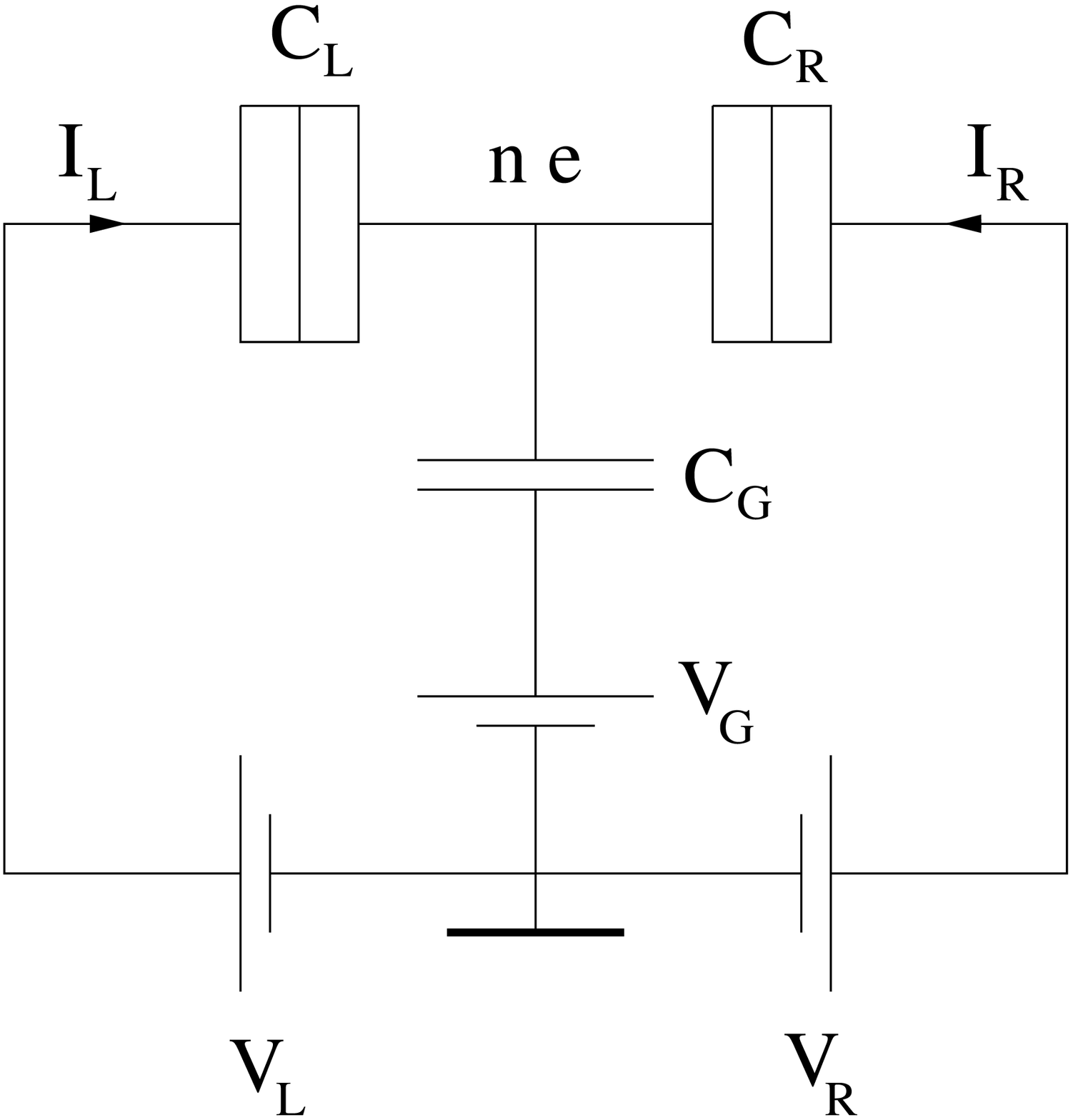,height=6.5cm}}
\caption{Single-electron transistor.}
\label{fig1}
\end{figure}

The single-electron transistor (see Fig.~\ref{fig1}) is modeled by the 
Hamiltonian
\begin{equation}
        H=H_{\rm L}+H_{\rm R}+H_{\rm I}+H_{\rm ch}+H_{\rm T}=H_0+H_{\rm T}.
\end{equation}
Here $H_{\rm r}=\sum_{kn}\epsilon^{\rm r}_{kn}a^\dagger_{{\rm r}kn}
a_{{\rm r}kn}$ and $H_{\rm I}=\sum_{qn}\epsilon_{qn} c^\dagger_{qn} c_{qn}$
describe the noninteracting electrons in the two leads r=L,R and on
the island. 
The index $n = 1, \dots N$ is the transverse channel index which includes the 
spin while the wave vectors $k$ and $q$ numerate the states of the electrons
within one channel.
In the following, we consider ``wide'' metallic junctions with $N \gg 1$.
The Coulomb interaction of the electrons on the island is described by
the capacitance model $H_{\rm ch}=E_{\rm C}(\hat{n}-n_{\rm x})^2$,
where $E_{\rm C}=e^2/(2C)$ with $C=C_{\rm L}+C_{\rm R}+C_{\rm g}$. 
The excess particle number operator on the island is given by
$\hat{n}$. Furthermore, the 'external charge'
$en_{\rm x}=C_{\rm L} V_{\rm L}+C_{\rm R} V_{\rm R}+C_{\rm g} V_{\rm g}$, 
accounts for the applied gate and transport voltages.
The charge transfer processes are described by 
the tunneling Hamiltonian
\begin{equation}
        H_{\rm T}=\sum_{\rm r=L,R}\sum_{kqn} T^{{\rm r}n}_{kq}
        a^\dagger_{{\rm r}kn}c_{qn}{\rm e}^{-i\hat{\varphi}}+ {\rm
        h.c.} \; .
\end{equation}
The matrix elements $T^{{\rm r}n}_{kq}$ are considered independent of the 
states $k$ and $q$ and the channel $n$.
They are related to the tunneling resistances $R_{\rm T,r}$ of the left and 
right junction by $1/ R_{\rm T,r} =  (2\pi e^2/ \hbar) N
N_{\rm r}(0) N_I(0) |T^{{\rm r}n}|^2$, where $N_{\rm I/r}(0)$ are the density 
of states of the island/leads at the Fermi level.
The operator ${\rm e}^{\pm i\hat{\varphi}}$ keeps track of the change of the 
charge on the island by $\pm e$.

We proceed using the diagrammatic technique developed in 
Refs.~\onlinecite{SS,KSS1}.
The nonequilibrium time evolution of the charge degrees of freedom on the 
island is described by a density matrix, which we expand in $H_{\rm T}$.
The reservoirs are assumed to remain in thermal equilibrium (with 
electrochemical potential $\mu_{\rm r}$) and are traced out using Wick's 
theorem, such that the Fermion operators are contracted in pairs.
For a large number of channels $N$, the ``simple loop'' configurations dominate 
where the two operators in $a^\dagger_{{\rm r}kn}c_{qn}$ from one term 
$H_{\rm T}$ are contracted with the corresponding two operators 
$c^\dagger_{qn}a_{{\rm r}kn}$ from another term $H_{\rm T}$, while the 
contribution of more complicated configurations are small.

\begin{figure}
\centerline{\psfig{figure=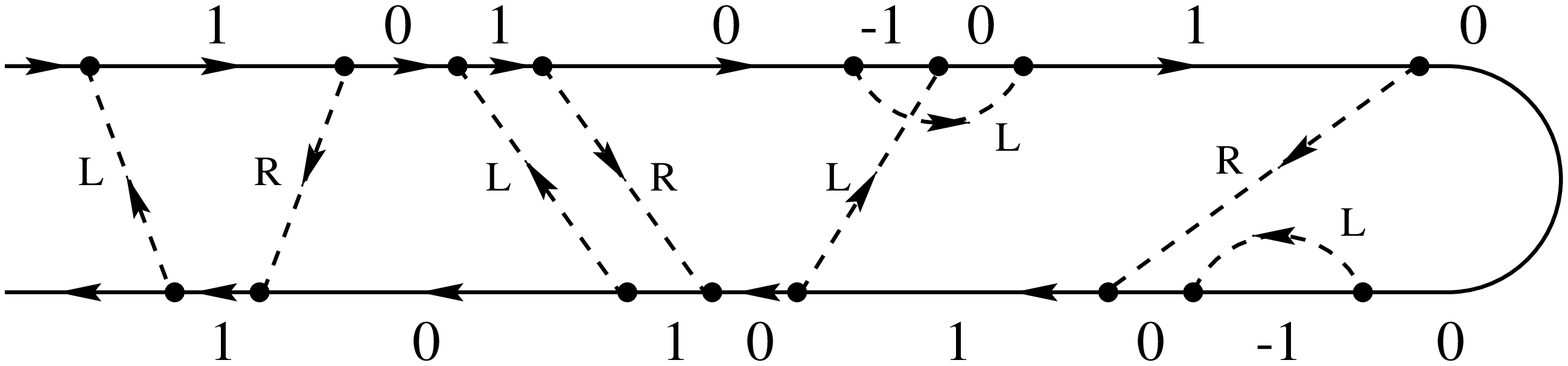,height=3cm}}
\caption{A diagram showing contributions to sequential tunneling 
        ($\Sigma_{0,1}^{(1)}$ and $\Sigma_{1,0}^{(1)}$) and 
        cotunneling: usual cotunneling with two electrons tunneling 
        coherently (contribution to $\Sigma_{0,0}^{(2)}$), vertex correction 
        ($\Sigma_{0,1}^{(2)}$), and propagator renormalization 
        ($\Sigma_{1,0}^{(2)}$).}
\label{fig2}
\end{figure}

The time evolution of the reduced density matrix in a basis of charge states is 
visualized in Fig.~(\ref{fig2}).
The forward and backward propagator (on the Keldysh contour) are coupled by 
``tunneling lines'', representing tunneling in junction ${\rm r}$. 
In Fourier space they are given by
\begin{equation}
        \alpha^\pm_{\rm r}(\omega) = \pm \alpha_0^{\rm r} {\omega-\mu_{\rm r} 
        \over \exp[\pm \beta(\omega-\mu_{\rm r})]-1} \, .
\end{equation}
if the line is directed backward/forward with respect to the closed time path.
Furthermore, we associate with each tunneling vertex at time $t$ a factor 
$\exp{(\pm i \Delta_n\,t)}$ depending on the energy difference of the adjacent 
charge states 
\begin{equation}
  \Delta_n=E_{\rm ch}(n+1)-E_{\rm ch}(n)=E_C \left[ 1+ 2 (n-n_x) \right].
\end{equation}
If the vertex lies on the backward path it acquires a factor $-1$. 
We define $\alpha_0=\sum_{\rm r} \alpha_0^{\rm r}$ and 
$\alpha (\omega)=\sum_{\rm r} \alpha_{\rm r} (\omega)$.

The time evolution of the system is determined by an infinite sequence of 
transitions between different charge states (see Fig.~\ref{fig2}).
This leads to a stationary probability distribution $p_n$ at the end.
As shown in Refs.~\onlinecite{SS,KSS1} the probabilities $p_n$ are described by
a master equation which reads in the stationary limit
\begin{equation}\label{master sigma}
        0 = \sum\limits_{n'\neq n}\left[ 
        p_{n'} \Sigma_{n',n} - p_n \Sigma_{n,n'} \right] \, . 
\end{equation}
Here $\Sigma_{n,n'}$ denotes the transition rate between $n$ and $n'$.
Only the latest transition, i.e., the rightmost irreducible part in the diagrams,
is written explicitly.
(We call a diagram part irreducible if there is no possible vertical cut 
separating the diagram into two parts.)
The sequence of the preceding transitions is summarized in the stationary
probability $p_n$ of charge state $n$.

\begin{figure}
\centerline{\psfig{figure=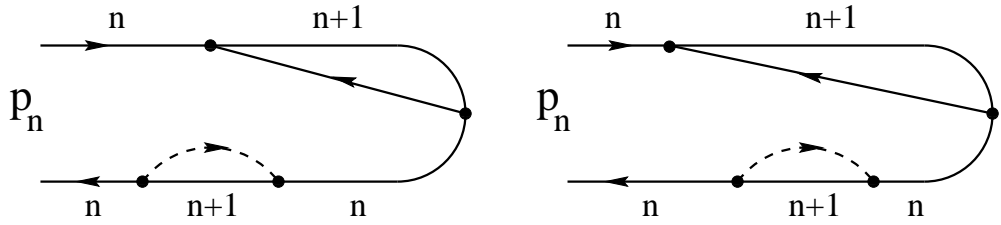,height=2.8cm}}
\caption{Two diagrams contributing to $C^{>(0,1)}(\omega,n)$.}
\label{fig3}
\end{figure}

The tunneling current $I_{\rm r}$ flowing into reservoir r can be expressed in
terms of the correlation functions for the island charge
\begin{eqnarray}
        C^>(t,t')&=&-i\langle e^{-i\varphi(t)} e^{i\varphi(t')} \rangle \, , 
        \label{correlation >} \\ \label{correlation <}
        C^<(t,t')&=&i\langle e^{i\varphi(t')} e^{-i\varphi(t)} \rangle \, .
\end{eqnarray}
Since the system is time-translational invariant, the correlation functions
depend only on the time difference $C(t,t')=C(t-t')$, and we can define
the Fourier transforms $C(\omega)=\int dt e^{i\omega t} C(t)$.
As shown in Refs.~\onlinecite{SS,KSS1} the tunneling current is
\begin{equation}\label{current}
        I_{\rm r}=-ie \int d\omega \left\{ \alpha_{\rm r}^+(\omega) C^>(\omega) +
                  \alpha_{\rm r}^-(\omega) C^<(\omega) \right\} \, .
\end{equation}
For each contribution to the correlation functions the corresponding diagram 
contains two external vertices (representing the $e^{\pm i\varphi}$ terms in
Eqs.~(\ref{correlation >}) and (\ref{correlation <}) ) connected by a solid 
line.
Two examples are shown in Fig.~\ref{fig3}.
The value of the diagram is given by the probability of the states indicated
on the left hand side of the diagram ($p_n$ in Fig.~\ref{fig3}) multiplied
with the value of the irreducible diagram part (consisting of two lines in 
Fig.~\ref{fig3}).
The probability includes the sequence of preceding transitions which are 
disconnected from the latest irreducible part.
While the calculation of the irreducible part follows straightforward from
the diagram rules, we need some extra relation in order to determine the 
stationary probabilities $p_n$.
This relation can be the master equation Eq.~(\ref{master sigma}).
But instead of Eq.~(\ref{master sigma}) we use an alternative condition which
is equivalent to the master equation.

For this task we introduce the quantities $C(\omega,n)$, $I_{\rm r}(n)$ and 
$I(n)=\sum_{\rm r}I_{\rm r}(n)$ as the contribution to the correlation function 
and current in which the rightmost vertex describes a transition from $n$ to 
$n+1$ or vice versa.
We can prove current conservation not only for the 
total current $\sum_{\rm r} I_{\rm r} = 0$ but also for each part $I(n)=0$
\cite{com}, i.e.,
\begin{equation}\label{master}
        \int d\omega \left\{ \alpha^+(\omega) C^>(\omega,n) +
                  \alpha^-(\omega) C^<(\omega,n) \right\} = 0\, .
\end{equation}
The normalization condition $\sum_n p_n = 1$ together with Eq.~(\ref{master}) 
determine the probabilities $p_n$ which can, then, be inserted in 
Eq.~(\ref{current}) to calculate the current.

\section{Systematic perturbation expansion}\label{section expansion}

For a systematic perturbation theory in powers of $\alpha_0$ we write
\begin{equation}
        p_n=\sum_{k=0}^\infty p_n^{(k)}
\end{equation}
for the probabilities, and
\begin{eqnarray}\label{expand C}
        C(\omega,n)=\sum_{k=0}^\infty C^{(k)}(\omega,n)
        =\sum_{k=0}^\infty \sum_{l=0}^k C^{(l,k-l)}(\omega,n)
\end{eqnarray}
for the correlation functions.
The index $k$ in $p_n^{(k)}$ and $C^{(k)}$ denotes the term of the order 
$\alpha_0^k$ in the expansion.
The term $C^{(l,k-l)}$ is the contribution to the correlation function $C^{(k)}$ 
of the order $k$ that is built up by a probability of the order $l$ and an
irreducible diagram part of the order $k-l$ with $0\le l \le k$.
(One may write $C^{(l,k-l)} \sim p^{(l)} \alpha^\pm (\omega_1) \ldots 
\alpha^\pm (\omega_{k-l})$ symbolically.)

The conservation rule Eq.~(\ref{master}) must hold in each order.
This allows us to determine the probabilities recursively.
If we know the probabilities $p_n^{(l)}$ of the order $l \le k-1$,
then we get $p_n^{(k)}$ from the $k+1$-th order term of the normalization
condition $\sum_n p_n^{(k)} = \delta_{k,0}$ and Eq.~(\ref{master}),
\begin{eqnarray}\label{probabilities}
        p_n^{(k)} \alpha^+(\Delta_n) - p_{n+1}^{(k)} \alpha^-(\Delta_n) 
        = \nonumber \\
        {1\over 2\pi i}\sum_{l=0}^{k-1} \int d\omega \left[
        \alpha^+ (\omega) C^{>(l,k-l)}(\omega,n) +
        \alpha^- (\omega) C^{<(l,k-l)}(\omega,n) \right] \, .
\end{eqnarray}
Here, we have used
\begin{eqnarray}
        C^{<(k,0)}(\omega,n)&=&2\pi i \,p_{n+1}^{(k)} \,\delta(\omega-\Delta_n)
        \label{C<k} \\ \label{C>k}
        C^{>(k,0)}(\omega,n)&=&- 2\pi i \, p_n^{(k)} \,\delta(\omega-\Delta_n) 
        \, .
\end{eqnarray}

Having determined the probabilities up to $p_n^{(k)}$ we get for the $k+1$-th 
term of the current Eq.~(\ref{current}) 
\begin{equation}\label{current k+1}
        I_{\rm r}^{(k+1)}=-ie \int d\omega 
        \left\{ \alpha_{\rm r}^+(\omega) C^{>(k)}(\omega) +
                  \alpha_{\rm r}^-(\omega) C^{<(k)}(\omega) \right\} \, .
\end{equation}

In lowest order (sequential tunneling), we have $k=0$ in 
Eq.~(\ref{probabilities}), i.e., 
$p_n^{(0)} \alpha^+(\Delta_n) - p_{n+1}^{(0)} \alpha^-(\Delta_n) = 0$ with
$\sum_n p_n^{(0)}=1$.
The solution in equilibrium is $p_n^{(0)}=\exp [-\beta E_{\rm ch}(n)]/Z$ with
$Z=\sum_n \exp [-\beta E_{\rm ch}(n)]$.
At low temperature (and voltage) at most two charge states 
($n=0,1$) are important, and all other states are exponentially suppressed.
Higher-order processes modify the occupation, and the probabilities for the 
other charge states are only algebraically suppressed, but not exponentially. 
This leads to correction terms $\langle n \rangle^{(k)}$ in the average 
occupation 
$\langle n \rangle = \sum_n n p_n= \sum_{k=0}^\infty \langle n \rangle^{(k)}$.

The current $I=I_{\rm L}=-I_{\rm R}$ follows from Eq.~(\ref{current k+1}) and is 
given in lowest order by
\begin{eqnarray}
        I^{(1)}={4 \pi^2 e\over h}\sum_n
        \left[p_n^{(0)}+p_{n+1}^{(0)}\right]    
        {\alpha_{\rm L}(\Delta_n) \alpha_{\rm R}(\Delta_n) \over 
        \alpha(\Delta_n)}
        \left[ f_{\rm R}(\Delta_n) - f_{\rm L}(\Delta_n) \right] \, ,
\end{eqnarray}
in which $f_{\rm r} (\omega) = 1 / \{ \exp [\beta (\omega-\mu_{\rm r} ) ] +1 \}$
is the Fermi function of reservoir r.
Higher-order terms account for new processes in which more electrons are involved
as well as renormalization effects of the dimensionless conductance and of the 
energy gap of adjacent charge states.

The rest of the paper is devoted to the calculation of the next order 
contribution (cotunneling), i.e., $k=1$ in Eqs.~(\ref{probabilities}) and 
(\ref{current k+1}).
We first calculate the correlation functions $C^{(0,1)}(\omega,n)$, insert the
result into Eq.~(\ref{probabilities}) to get the corrections to the
probabilities $p_n^{(1)}$.
Then we have $C^{(1,0)}(\omega,n)$ and $C^{(1)}(\omega)$ from Eqs.~(\ref{C<k}), 
(\ref{C>k}) and (\ref{expand C}), and the current $I^{(2)}$ folloes from
Eq.~(\ref{current k+1}).

\section{General results}\label{section general}

The first task is to evaluate all diagrams contributing to the correlation 
functions $C^{(0,1)}(\omega,n)$.
For $C^<$ and $C^>$ there are 32 diagrams each.
They contain one tunnel line and two external vertices.
In Fig.~\ref{fig3} we show two such diagrams which contribute to 
$C^{>(0,1)}(\omega,n)$.
The other diagrams are generated from these two by putting the right and/or the 
left vertex of the tunnel line on the opposite propagator and/or change the 
direction of the tunneling line.
In each case we have to adjust the charge states at the propagators in such a way
that the rightmost (external) vertex describes a transition from $n+1$ to $n$
(or vice versa).
By this procedure, we get in total 16 terms. 
The other 16 terms are obtained by mirroring the old diagrams with respect to 
a horizontal line and changing the direction of all lines.
The diagrammatic rules imply that the mirrored diagrams are minus the complex 
conjugate of the old ones.

The first and second diagram in Fig.~\ref{fig3} have the values
\begin{equation}
         p_n^{(0)} \int d\omega' \, \alpha^+(\omega') 
        \, {1\over \Delta_n - \omega' + i0^+}
        \, {1\over \omega - \omega' + i0^+}
        \, {1\over \omega -\Delta_n + i0^+}
\end{equation}
and
\begin{equation}
        p_n^{(0)} \int d\omega' \, \alpha^+(\omega') 
        \, {1\over \omega - \omega' + i0^+}
        \, {1\over \left( \omega -\Delta_n + i0^+ \right) ^2} \, .
\end{equation}
The integrands have poles.
But these poles are regularized in a natural way (this follows from our theory 
and has {\it not} to be added by hand) as Cauchy's principal values and delta 
functions,
$1 / (x+i0^+) = P [1/ x] - i \pi \delta(x)$, and their derivative 
$1 / (x+i0^+)^2 = -(d/ dx) P [1/x] + i \pi \delta'(x)$.
The integrals are well-defined at these points.
Another condition for the existence of the integrals concerns the behaviour at 
$|\omega'| \rightarrow \infty$.
Since $\alpha (\omega')$ raises linearly at large energies, the integrands
of some integrals may not decay fast enough at $|\omega'| \rightarrow
\infty$ to ensure convergence for a single diagram.
The sum of all diagrams, however, is convergent.
To see this, we use for each diagram a Lorentzian cutoff $U^2/(U^2+\omega'^2)$ 
with $U$ larger than all relevant energy scales.
Then, all intermediate results become convergent but may be $U$-dependent.
To be more precisely, the cutoff enters only in the form
$W = \ln \left( \beta U \over 2 \pi \right)$.
While the $W$ is present at some intermediate steps (even in the probabilities
$p_n^{(1)}$), we find that the physical quantities like the average charge and 
the current are independent of $W$.
This means that the divergencies are only an artefact of considering single
diagrams and are not present for the sum of all contributions.
 
Defining $\phi_{n,{\rm r}}=\alpha_0^{\rm r} (\Delta_n -\mu_{\rm r}) \, {\rm Re} 
\, \Psi \left(i{\beta\over 2\pi} (\Delta_n -\mu_{\rm r})\right)$ and 
$\phi_n=\sum_{\rm r} \phi_{n,{\rm r}}$, where $\Psi$ denotes the digamma 
function, and making use of $\Delta_{n+1}-\Delta_n=2E_{\rm C}$, we obtain
the lengthy but complete result
\begin{eqnarray}\label{correlation 1}
        &&C^{>(0,1)}(\omega,n)=2\pi i
        \nonumber \\ &&
        \left\{ - \left[ p_{n-1}^{(0)} \alpha^+(\Delta_n+\Delta_{n-1}-\omega)
                      + p_n^{(0)} \alpha^-(\omega) \right]
        {\rm Re} \left[ {1\over \omega -\Delta_n+i0^+} R_n(\omega) \right]
        \right. \nonumber \\ && \left. + 
        \left[ p_n^{(0)} \alpha^+(\Delta_{n+1}+\Delta_n-\omega)
                      + p_{n+1}^{(0)} \alpha^-(\omega) \right]
        {\rm Re} \left[ {1\over \omega -\Delta_n+i0^+} R_{n+1}(\omega) \right]
        \right. \nonumber \\ && \left. +
        p_{n-1}^{(0)} \delta (\omega-\Delta_{n-1}) 
        {\phi_n-\phi_{n-1}\over 2E_{\rm C}}
        +
        p_n^{(0)} \delta (\omega-\Delta_n) 
        {\phi_{n+1}-\phi_{n-1}\over 2E_{\rm C}}
        +
        p_{n+1}^{(0)} \delta (\omega-\Delta_{n+1}) 
        {\phi_{n+1}-\phi_n\over 2E_{\rm C}}
        \right. \nonumber \\ && \left. 
        - \delta (\omega-\Delta_n) \left[
        p_{n-1}^{(0)} \partial \phi_{n-1} +
        p_n^{(0)} \partial (\phi_n + \phi_{n+1}) +
        p_{n+1}^{(0)} \partial \phi_n \right]
        +
        p_n^{(0)} \delta' (\omega-\Delta_n) \left[ 
        2 \phi_{n} - \phi_{n+1} - \phi_{n-1} \right] 
        \right. \nonumber \\ && \left. +
        \alpha_0 W p_{n-1}^{(0)} \left[ 
        \delta (\omega-\Delta_n) - \delta (\omega-\Delta_{n-1}) \right]
        + \alpha_0 W p_{n+1}^{(0)} \left[ 
        \delta (\omega-\Delta_n) - \delta (\omega-\Delta_{n+1}) \right]
        \right\} \, .
\end{eqnarray}
Here $\partial$ stands for $-{1\over 2E_C}{\partial \over \partial n_x}$ and
we have used the abbreviation
\begin{equation}
        R_n(\omega)={1\over \omega-\Delta_n+i0^+} - 
        {1\over \omega-\Delta_{n-1}+i0^+} \, .
\end{equation}

The other contribution to 
$C^{>(1)}(\omega,n)=C^{>(0,1)}(\omega,n)+C^{>(1,0)}(\omega,n)$ involves 
probabilities in higher order, see Eq.~(\ref{C>k}) for $k=1$.

The expression for $C^{<(1)}(\omega,n)$ looks very similar.
It is obtained from $C^{>(1)}(\omega,n)$ by changing the direction of all
lines and adjusting the charge states.
This implies $\alpha^+ \leftrightarrow \alpha^-$, 
$\Delta_{n-1} \leftrightarrow \Delta_{n+1}$,
$\phi_{n-1} \leftrightarrow \phi_{n+1}$,
$2E_{\rm C} \leftrightarrow -2E_{\rm C}$, 
$p_{n-1} \leftrightarrow p_{n+2}$, $p_n \leftrightarrow p_{n+1}$, 
$R_n \leftrightarrow -R_{n+1}$,
and, furthermore, there is a global minus sign.

We can solve Eq.~(\ref{probabilities}) with the help of $C^{>(0,1)}(\omega,n)$ 
and $C^{<(0,1)}(\omega,n)$ for the probabilities $p_n^{(1)}$. 
Due to the last line of Eq.~(\ref{correlation 1}) the corrections of the 
probabilities, $p_n^{(1)}$, are $W$-dependent.
But the $W$-dependence is of the form $-\alpha_0 W \left[ 2p_n^{(0)} - 
p_{n+1}^{(0)} - p_{n-1}^{(0)} \right]$.
As a consequence, in $C^{(1)}(\omega)=\sum_n C^{(1)}(\omega,n)$ as well as in
$\langle n \rangle^{(1)}$ the $W$-independent terms drop out, i.e., there is no 
divergency problem for the physical quantities.

Now we are ready to formulate the most important result of this paper.
Inserting the result for $C^{(1)}(\omega)$ in Eq.~(\ref{current k+1}) with
$k=1$ we obtain the cotunneling current $I^{(2)}$ which can be divided into
four parts, $I^{(2)} = \sum_{i=1}^4 I^{(2)}_i $ with
\begin{eqnarray}
        \label{general 1}
        I^{(2)}_1&=&
        \sum_n p_n^{(0)} \int d\omega \, B(\omega) \, \alpha(\omega) \,
        {\rm Re} \, R_n(\omega)^2 \, ,
        \\ \label{general 2}
        I^{(2)}_2 &=& - {1\over 2}\sum_n \left( p_n^{(0)}+p_{n+1}^{(0)}\right)
        B(\Delta_n) \int d\omega \, \alpha(\omega) {\rm Re} 
        \left[ R_n(\omega)^2 + R_{n+1}(\omega)^2 \right] 
        \nonumber \\ &&
        + \sum_n \left( \hat{p}_n+\hat{p}_{n+1}\right) B(\Delta_n) \, ,
        \\ \label{general 3}
        I^{(2)}_3 &=& - {1\over 2}
        \sum_n  \left( p_n^{(0)}+p_{n+1}^{(0)}\right) 
        {\partial B(\Delta_n) \over \partial \Delta_n} \, 
        \int d\omega \, \alpha(\omega) {\rm Re} \, 
        \left[ R_n(\omega) - R_{n+1}(\omega) \right] \, ,
        \\ \label{general 4}
        I^{(2)}_4&=& {4\pi^2e \over h} \sum_n \int d\omega \,
        \left\{ 
         - p_{n-1}^{(0)} \alpha^+(\Delta_n+\Delta_{n-1}-\omega) 
        { \alpha_{\rm L}^+(\omega) \alpha_R(\Delta_n) 
           \over \alpha (\Delta_n) }
        {\rm Re} \, \left[ {1\over \omega-\Delta_n+i0^+}   R_n(\omega) \right]
        \right. \nonumber \\ && \left. \qquad\qquad\qquad
         + p_{n+1}^{(0)} \alpha^-(\Delta_n+\Delta_{n-1}-\omega) 
        { \alpha_{\rm L}^-(\omega) \alpha_R(\Delta_n) 
           \over \alpha (\Delta_n) }
        {\rm Re} \, \left[ {1\over \omega-\Delta_n+i0^+}   R_n(\omega) \right]
        \right. \nonumber \\ && \left. \qquad\qquad\qquad
         + p_n^{(0)} \alpha^+(\Delta_{n+1}+\Delta_n-\omega) 
        { \alpha_{\rm L}^+(\omega) \alpha_R(\Delta_n) 
           \over \alpha (\Delta_n) }
        {\rm Re} \, \left[ {1\over \omega-\Delta_n+i0^+} R_{n+1}(\omega) \right]
        \right. \nonumber \\ && \left. \qquad\qquad\qquad
         - p_{n+2}^{(0)} \alpha^-(\Delta_{n+1}+\Delta_n-\omega) 
        { \alpha_{\rm L}^-(\omega) \alpha_R(\Delta_n) 
           \over \alpha (\Delta_n) }
        {\rm Re} \, \left[ {1\over \omega-\Delta_n+i0^+} R_{n+1}(\omega) \right]
        \right\}
        \nonumber \\ &&
        \qquad\qquad\qquad - \, ({\rm L} \leftrightarrow {\rm R}) \, .
\end{eqnarray}
Here we used the definitions
\begin{equation}
  B(\omega) \equiv {4 \pi^2 e\over h}
  {\alpha_{\rm L}(\omega) \alpha_{\rm R}(\omega) \over 
    \alpha(\omega)}
  \left[ f_{\rm R}(\omega) - f_{\rm L}(\omega) \right]
\end{equation}
and 
\begin{equation}
  \hat{p}_n \equiv p_n^{(1)} - \alpha_0 W \left[ 2p_n^{(0)} - p_{n+1}^{(0)} - 
    p_{n-1}^{(0)} \right] - (p_n^{(0)} -  p_{n+1}^{(0)}) \partial \phi_{n-1} 
  + (p_{n+1}^{(0)} -  p_n^{(0)}) \partial \phi_n \, .
\end{equation}
In the low temperature limit, $B(\Delta_0)$ is the sequential tunneling result.
It enters in $I^{(2)}_1,I^{(2)}_2$, and $I^{(2)}_3$ in different ways.

The first term $I_1^{(2)}$ describes processes in which one electron enters
and other one leaves the island coherently.
In the Coulomb blockade regime only one charge state is occupied in lowest order
at low temperature, $p_n^{(0)} = \delta_{n,0}$.
At $T=0$, the integrand is zero at the poles of $R_n(\omega)$, and we can omit
the term $i0^+$.
In this case, we recover from Eq.~(\ref{general 1}) the usual ''cotunneling'' 
result \cite{Ave-Naz}.
At finite temperature, however, the term  $+i0^+$ is needed for regularization 
which is not provided by previous theories \cite{com3}. 
The result Eq.~(\ref{general 1}) is also well-defined for $T \neq 0$. 

Furthermore, we are able to describe the system at resonance.
In this regime, $I^{(2)}_2$ and $I^{(2)}_3$ become important.
These terms describe the renormalization of the tunneling conductance 
$\alpha_0$ and the energy gap $\Delta_0$, respectively.
They can be traced back to vertex correction and propagator renormalization 
diagrams (see Fig.~\ref{fig2}).
We will discuss these contributions in more detail in Sec.\ref{section low temp}
and Sec.\ref{section rg}.

At temperature or bias voltage, there are processes in which two electrons
enter or leave the island coherently.
After such a process, the island charge has increased or decreased by $2e$.
These processes are described by the fourth term $I_4^{(2)}$, 
Eq.~(\ref{general 4}).

All integrals in Eqs.~(\ref{general 1}) - (\ref{general 4}) can be performed
analytically.
We do not write down the lengthy results for the general case here.
They can be found for the most interesting limits in Sec.~\ref{section low temp}
and \ref{section all temp}.
The quantities $\hat{p}_n$ which enter Eq.~(\ref{general 2}) are determined 
by $\sum_n \hat{p}_n =0$ and
\begin{eqnarray}\label{probabilities hat}
  &&\hat{p}_n \alpha^+(\Delta_n) - \hat{p}_{n+1} \alpha^-(\Delta_n) =  
  \nonumber \\ 
  &&\sum_{r,r'} {p_{n-1}^{(0)} - e^ { \beta (\Delta_n+\Delta_{n-1} - \mu_r
      -\mu_{r'}) } p_{n+1}^{(0)} \over e^{ \beta (\Delta_n+\Delta_{n-1} 
      - \mu_r -\mu_{r'}) } -1 } \, \alpha_0^r
  \nonumber \\ 
  &&\times \left\{ - \partial \left[ (\Delta_n-\mu_r)\phi_{n-1,r'} 
      + ( \Delta_{n-1}-\mu_r )\phi_{n,r'} \right]
    - {\Delta_{n-1}-\mu_r \over E_C} \phi_{n-1,r'} 
    + {\Delta_n-\mu_r \over E_C} \phi_{n,r'}  \right\} 
  \nonumber \\ 
  && + \sum_{r,r'} {p_n^{(0)} - e^ { \beta (\Delta_{n+1}+\Delta_n - \mu_r
      -\mu_{r'}) } p_{n+2}^{(0)} \over e^{ \beta (\Delta_{n+1}+\Delta_n 
      - \mu_r -\mu_{r'}) } -1 } \, \alpha_0^r
  \nonumber \\ 
  && \times \left\{ - \partial \left[ (\Delta_{n+1}-\mu_r)\phi_{n,r'} 
      + ( \Delta_n-\mu_r )\phi_{n+1,r'} \right]
    - {\Delta_n-\mu_r \over E_C} \phi_{n,r'} 
    + {\Delta_{n+1}-\mu_r \over E_C} \phi_{n+1,r'}  \right\} 
  \nonumber \\ 
  && - \left[ p_n^{(0)} \partial \alpha^+(\Delta_n)
    - p_{n+1}^{(0)} \partial \alpha^-(\Delta_n) \right]
  \left[ 2 \phi_n - \phi_{n+1} - \phi_{n-1} \right] 
\end{eqnarray}

\begin{figure}
\centerline{\psfig{figure=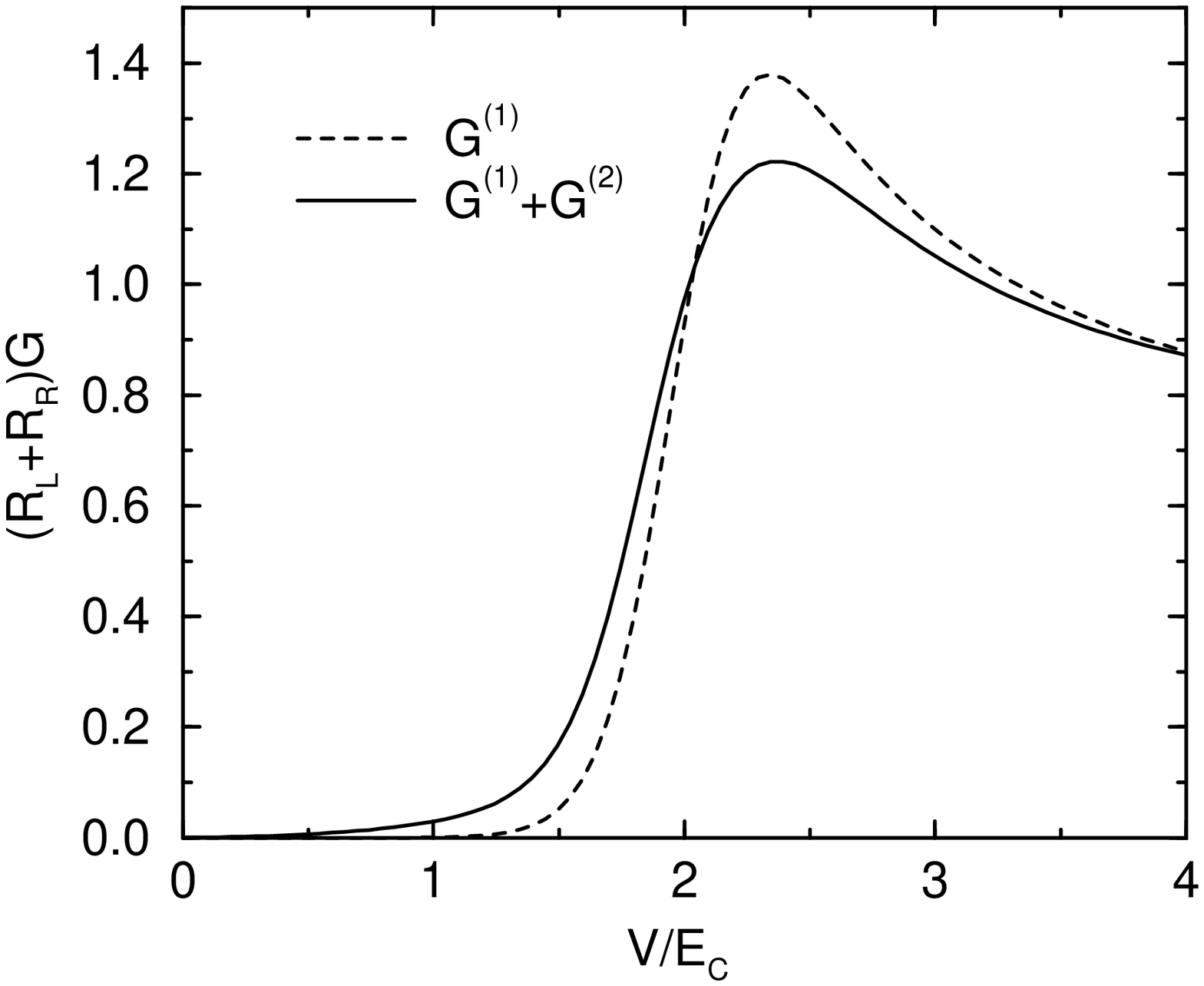,height=7cm}}
\caption{The differential conductance as a function of transport voltage
        for $T/E_C=0.05$, $n_x=0$, and $\alpha_0=0.01$:
        sequential tunneling and sequential plus cotunneling contribution.}
\label{fig4}
\end{figure}

\begin{figure}
\centerline{\psfig{figure=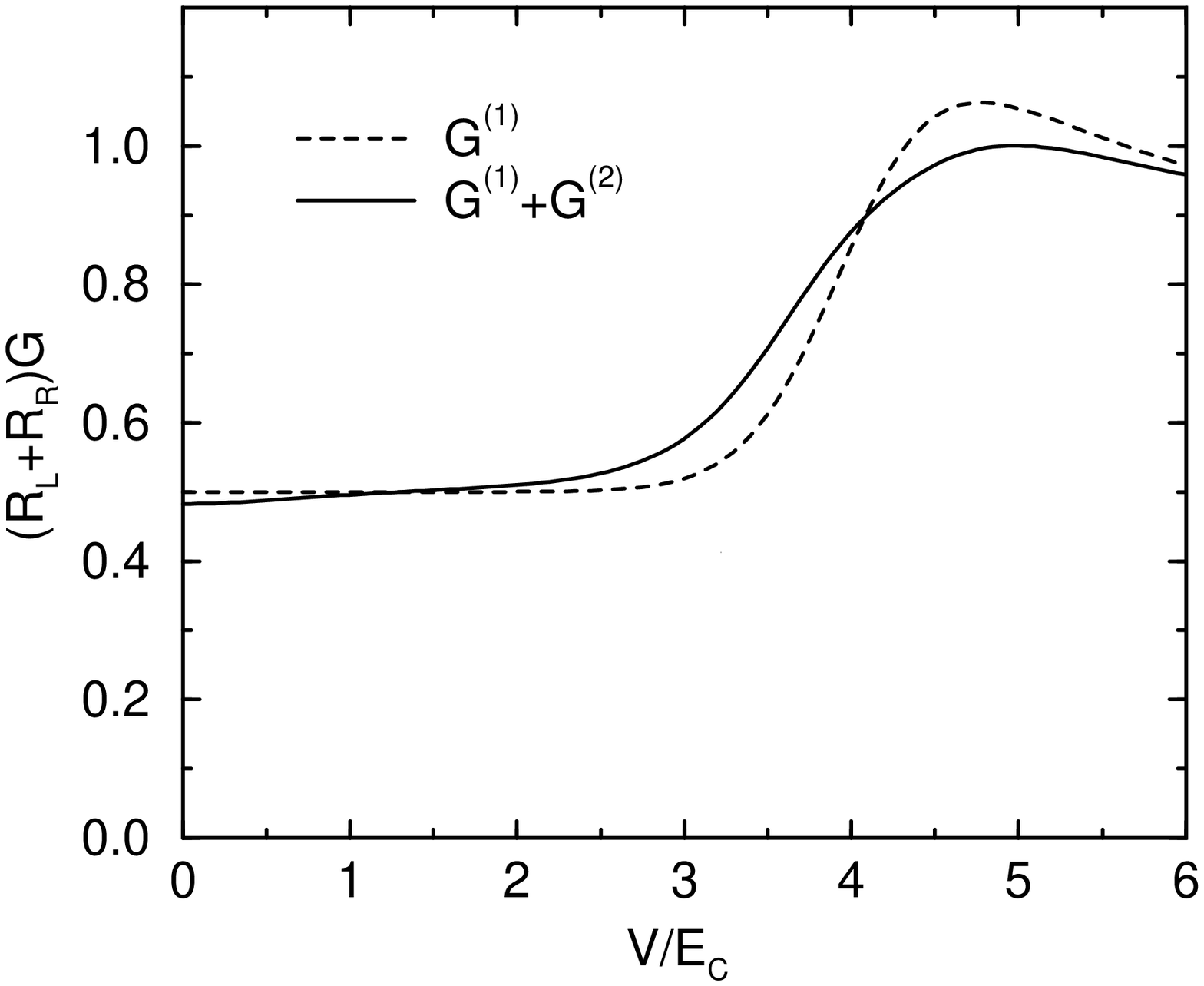,height=7cm}}
\caption{The differential conductance as a function of transport voltage
        for $T/E_C=0.1$, $n_x=0.5$, and $\alpha_0=0.01$:
        sequential tunneling and sequential plus cotunneling contribution.}
\label{fig5}
\end{figure}

In principle, the problem is now solved.
One can solve Eq.~(\ref{probabilities hat}) for $\hat{p}_n$ 
numerically and will then get the average charge and the current.
In Figs.~\ref{fig4} and \ref{fig5} we show the differential conductance as a
function of the transport voltage for two different values of the gate charge.
There are steps at voltages at which new charge states become important.
Due to cotunneling the steps are washed out.
Furthermore, there is a finite conductance in the Coulomb blockade regime 
although the sequential tunneling contribution is exponentially suppressed.

But we can do even better and find analytic expressions for the two most 
interesting cases. 
The first one covers the regime in which only two charge states are occupied in
lowest order (then four charge states have corrections in first order). 
In this regime the Coulomb oscillations are most pronounced and the system is 
expected to behave like a multichannel Kondo model.
But to study the crossover from the Coulomb blockade to the classical (high 
temperature) regime, one has to allow for all temperatures.
This is done in the second case, in which we consider the linear response.

\section{Low temperature nonequilibrium transport}\label{section low temp}

At low temperature, $k_BT \ll E_{\rm C}$, and a transport voltage which may be 
finite but small enough, only two charge states ($n=0,1$) are occupied in 
lowest order, $p_n^{(0)}\approx 0$ for $n\neq 0,1$.

Sequential tunneling gives for the average charge just 
$\langle n \rangle^{(0)}=p_1^{(0)}=\alpha^+(\Delta_0)/ \alpha(\Delta_0)$,
and the current is $I^{(1)}=B(\Delta_0)$.

The probabilities in next order, $p_n^{(1)}$, are determined by 
Eq.~(\ref{probabilities}).
In the low temperature limit, the first and second term in 
Eq.~(\ref{probabilities hat}) are exponentially small.
We find
\begin{eqnarray}
        p_{-1}^{(1)} &=& - p_0^{(0)} \partial \phi_{-1} + p_0^{(0)} \alpha_0 W
        \\
        p_0^{(1)} &=& p_0^{(0)} \partial \phi_{-1} 
        - (p_1^{(0)} - p_0^{(0)}) \partial \phi_0 + \partial p_0^{(0)} \left[
        2 \phi_0 - \phi_{-1} -\phi_1 \right]
        + (p_1^{(0)} - 2 p_0^{(0)}) \alpha_0 W
        \\
        p_1^{(1)} &=& (p_1^{(0)} - p_0^{(0)}) \partial \phi_0 
        + p_1^{(0)} \partial \phi_1 + \partial p_1^{(0)} \left[
        2 \phi_0 - \phi_{-1} -\phi_1 \right]
        + (p_0^{(0)} - 2 p_1^{(0)}) \alpha_0 W
        \\
        p_2^{(1)} &=& - p_1^{(0)} \partial \phi_1 + p_1^{(0)} \alpha_0 W
\end{eqnarray}
while all other corrections vanish, and therefore
\begin{equation}
        \langle n \rangle^{(1)}= -{1\over 2 E_{\rm C}}
        {\partial \over \partial n_{\rm x}}
        \left[ p_0^{(0)} \left( \phi_{-1} - \phi_0  \right)
             + p_1^{(0)} \left( \phi_0 - \phi_1  \right) \right]
\end{equation}

The total second order, ``cotunneling'' current 
Eqs.~(\ref{general 1}) - (\ref{general 4}) simplifies to
\begin{eqnarray}\label{current1}
        I^{(2)}_1 &=&
        \int d\omega \, B(\omega) \alpha(\omega) \,
        {\rm Re} \,  \left[ p_0^{(0)} R_0(\omega)^2 +
                            p_1^{(0)} R_1(\omega)^2 \right] \, ,
        \\ \label{current2}
        I^{(2)}_2 &=& - {1\over 2}B(\Delta_0)
        \int d\omega \, \alpha(\omega) \, {\rm Re}
        \left[ R_0(\omega)^2+R_1(\omega)^2 \right] \, ,
        \\ \label{current3}
        I^{(2)}_3 &=&
        - {1\over 2}{\partial B(\Delta_0) \over \partial \Delta_0} \, 
        \int d\omega \, \alpha(\omega) \, {\rm Re} 
        \left[ R_0(\omega)-R_1(\omega) \right] \, ,
\end{eqnarray}
and $I_4^{(2)} = 0$.
All integrals in Eqs.~(\ref{current1}) - (\ref{current3}) can be performed 
analytically.
The result is
\begin{eqnarray} \label{current1a}
        I^{(2)}_1 &=& 
        {4\pi^2 e\over h} \alpha_0^{\rm R} 
        \sum_{n=0,1} p_n^{(0)} \left\{ 
        {\Delta_{n-1} -\mu_{\rm R} \over E_{\rm C}} \phi_{n-1,{\rm L}} 
        - {\Delta_n -\mu_{\rm R} \over E_{\rm C}} \phi_{n,{\rm L}} 
        \right. \nonumber \\ && \left. \qquad \qquad \qquad     
        + \partial \left[
        (\Delta_{n-1} -\mu_{\rm R}) \phi_{n-1,{\rm L}} + 
        (\Delta_n -\mu_{\rm R}) \phi_{n,{\rm L}} \right] \vphantom{E\over E}
        \right\}  \, \,
        - \, ({\rm L} \leftrightarrow {\rm R}) \, ,
        \\ \label{current2a}
        I^{(2)}_2 &=&   B(\Delta_0) \left\{ 
        {\phi_{-1} - \phi_1\over E_{\rm C}}
        + \partial [2\phi_0+\phi_{-1}+\phi_1] \right\} \, ,
        \\ \label{current3a}
        I^{(2)}_3 &=& 
        \left[ {\partial B(\Delta_0) \over \partial \Delta_0} \right]
        (2\phi_0-\phi_{-1}-\phi_1) \, .
\end{eqnarray}

In the Coulomb blockade regime, we have $p_0^{(0)}=1$, $p_1^{(0)}=0$ and 
$I^{(1)}(\Delta_0)=0$, which yields $I^{(1)}_2=I^{(1)}_3=0$.
The only contribution is the term with $n=0$ in Eq.~(\ref{current1}). 
This gives the well-known result of inelastic cotunneling \cite{Ave-Naz}. 

At resonance ($p_0^{(0)}\neq 0 \neq p_1^{(0)}$), the terms describing
the renormalization of the tunneling conductance $\alpha_0$ and the energy 
gap $\Delta_0$, $I^{(2)}_2$ and $I^{(2)}_3$, become important.
Suppose that the current can be expressed by the sequential tunneling result 
with renormalized parameters $\tilde{\alpha}$ and $\tilde{\Delta}$ plus some 
regular terms,  $I(\alpha_0,\Delta_0) = I^{(1)}(\tilde{\alpha},\tilde{\Delta}) + 
{\rm regular \phantom{*} terms}$.
The expansion up to ${\cal O}(\alpha_0^2)$ of
\begin{equation}
        I^{(1)}(\tilde{\alpha},\tilde{\Delta}) =
        {\tilde{\alpha} - \alpha_0 \over \alpha_0} I^{(1)}(\alpha_0,\Delta_0) +
        (\tilde{\Delta} - \Delta_0 ) {\partial I^{(1)}(\alpha_0,\Delta_0) \over
        \partial \Delta_0} \, .
\end{equation}
and the identification with Eqs.~(\ref{current2a}) and (\ref{current3a}) yield 
the lowest-order corrections for the renormalized quantities.

\begin{figure}
\centerline{\psfig{figure=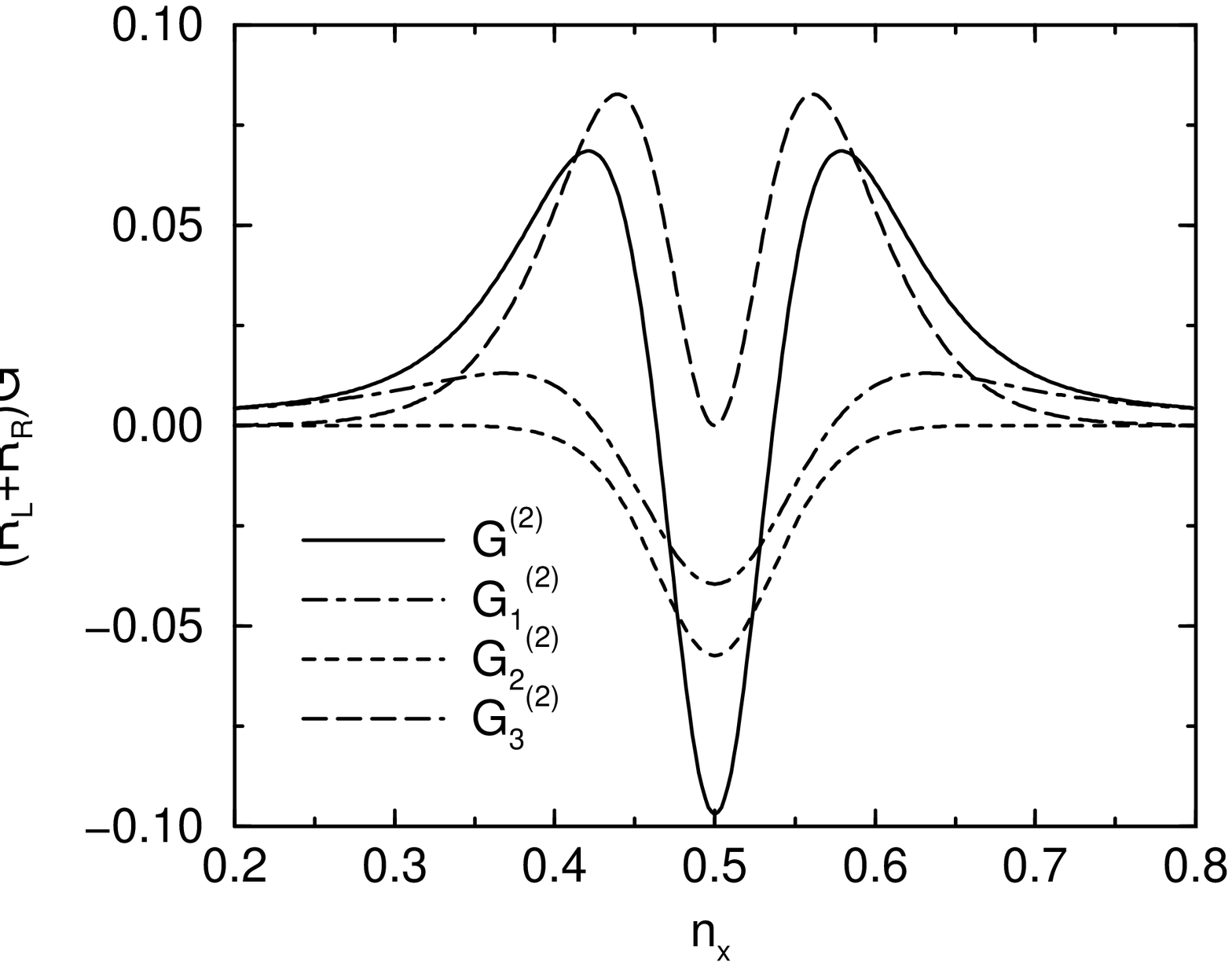,height=7cm}}
\caption{The second-order contribution of the differential conductance 
        $G^{(2)}=\sum_{i=1}^3 G_i^{(2)}$ for $T/E_{\rm C}=0.05$, 
        $\alpha_0=0.04$ and $V=0$.}
\label{fig6}
\end{figure}

\begin{figure}
\centerline{\psfig{figure=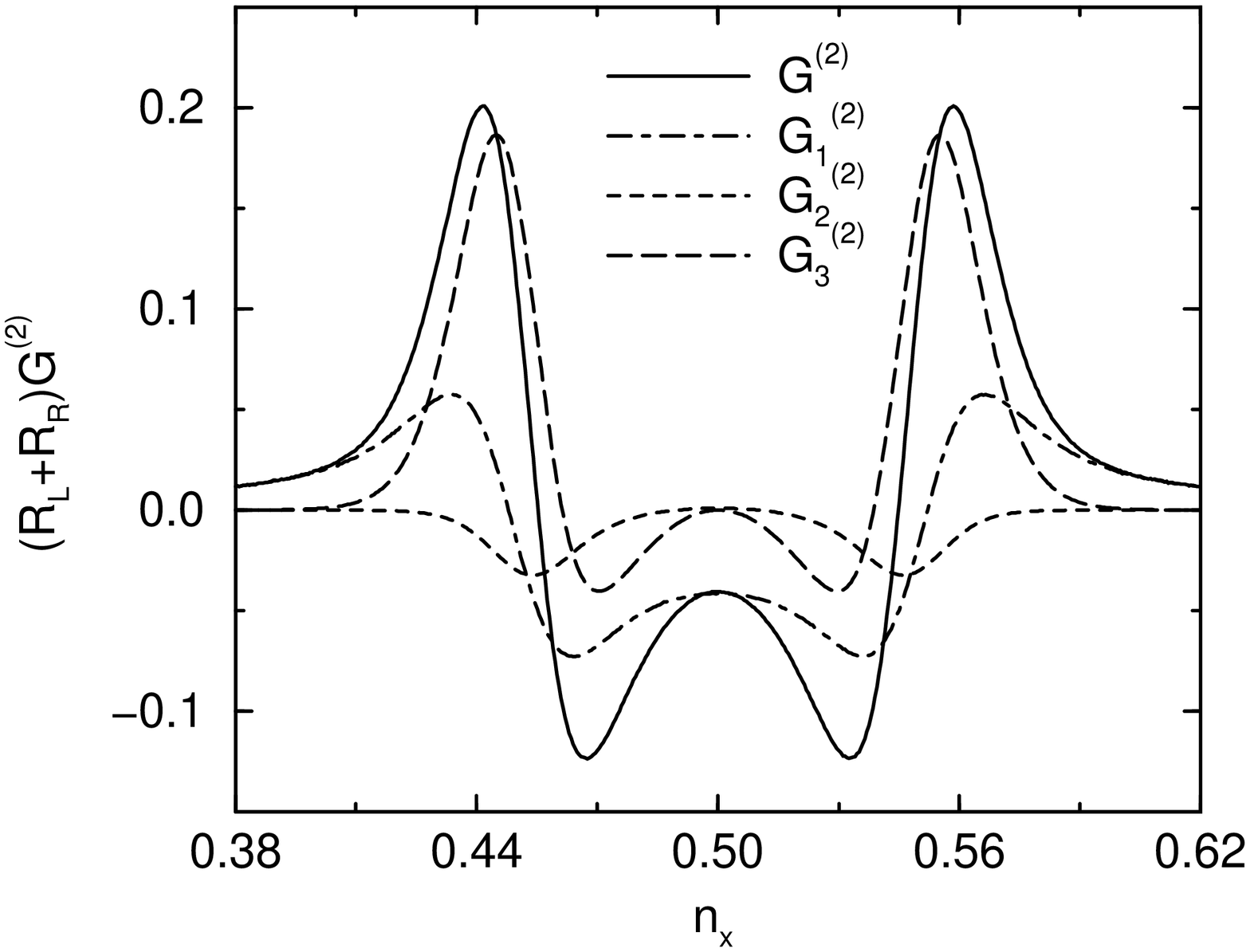,height=7cm}}
\caption{The second-order contribution to the differential conductance 
        $G^{(2)}=\sum_{i=1}^3 G_i^{(2)}$ for $T/E_{\rm C}=0.01$, 
        $\alpha_0=0.02$ and $V/E_{\rm C}=0.2$.}
\label{fig7}
\end{figure}

In Figs.~\ref{fig6} and \ref{fig7} we show the second-order contribution
to the differential conductance $G=\partial I/ \partial V$ for the linear
and nonlinear regime (in the following we choose 
$\alpha_0^{\rm L}=\alpha_0^{\rm R}$).
The dimensionless conductance $\tilde{\alpha}$ as well as energy gap 
$\tilde{\Delta}$ is decreased in comparison to the unrenormalized values
$\alpha_0$ and $\Delta_0$.
For this reason, $I_2^{(2)}$ is negative since the conductance is reduced,
and $I_3^{(2)}$ is positive since the system is effectively ''closer'' to the 
resonance.

\begin{figure}
\centerline{\psfig{figure=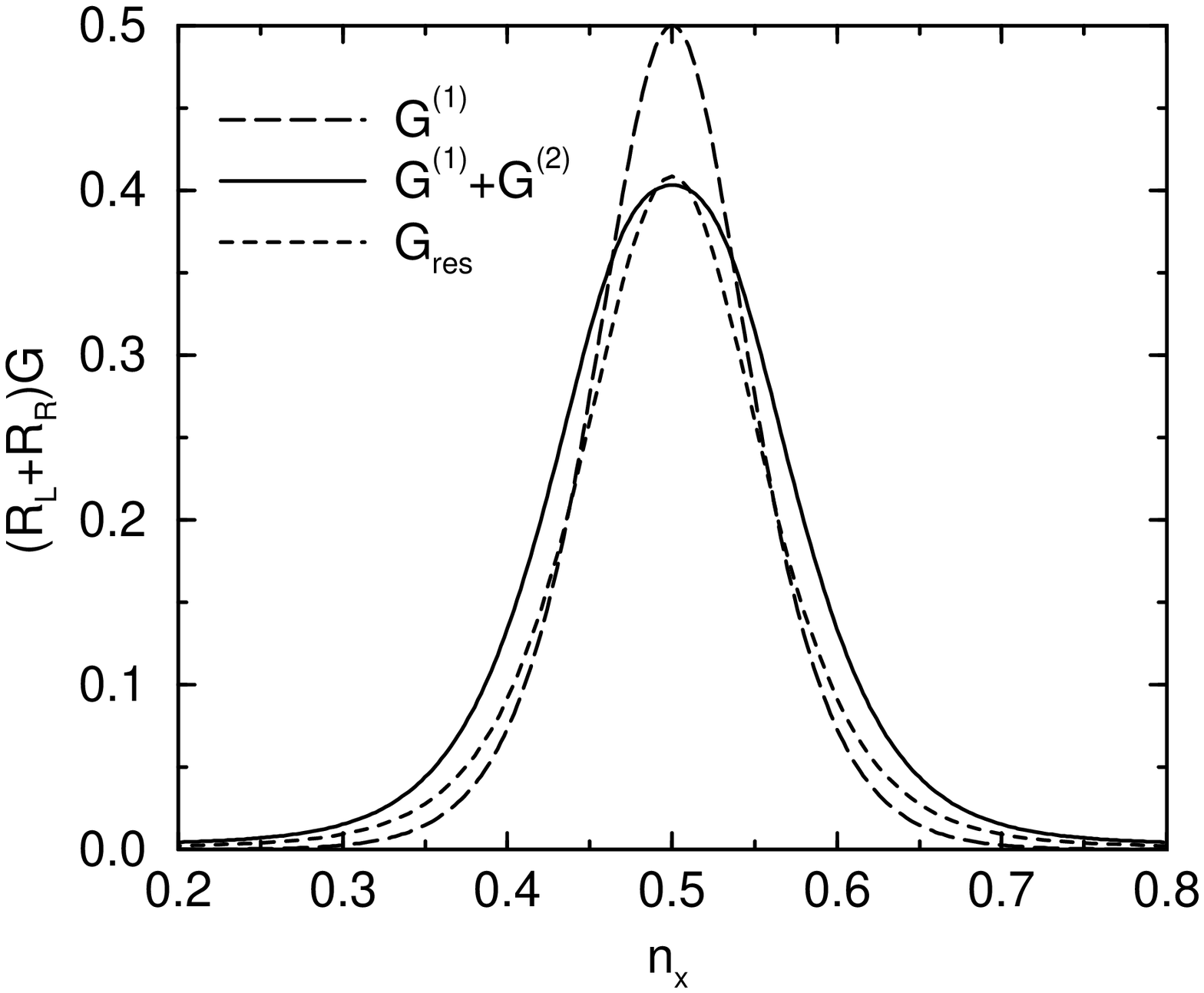,height=7cm}}
\caption{The differential conductance for $T/E_{\rm C}=0.05$, $\alpha_0=0.04$ 
        and $V=0$: sequential tunneling, sequential plus cotunneling, and 
        resonant tunneling approximation.}
\label{fig8}
\end{figure}

\begin{figure}
\centerline{\psfig{figure=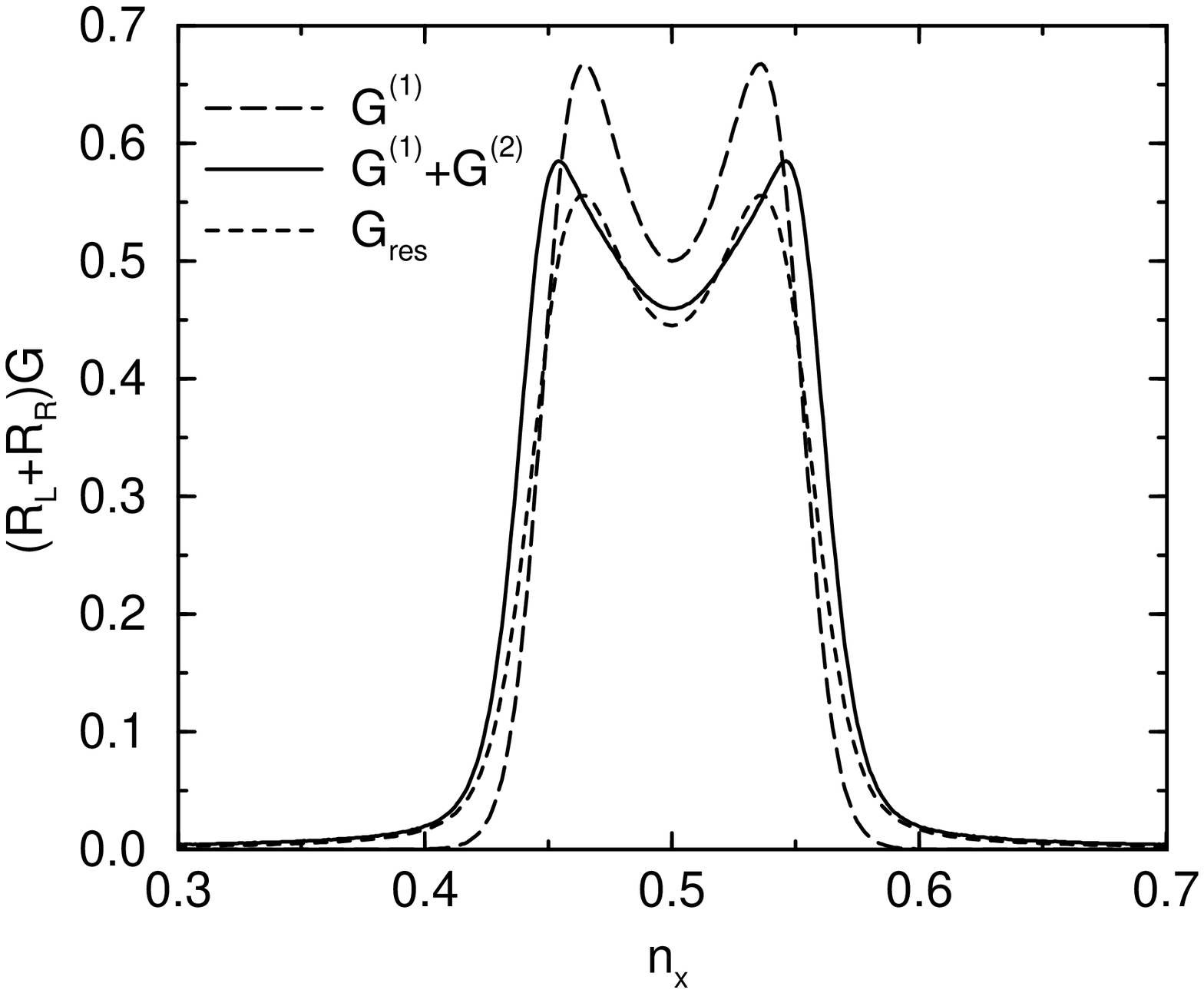,height=7cm}}
\caption{The differential conductance for $T/E_{\rm C}=0.01$, $\alpha_0=0.02$ 
        and $V/E_{\rm C}=0.2$: sequential tunneling, sequential plus 
        cotunneling, and resonant tunneling approximation.}
\label{fig9}
\end{figure}

In Figs.~\ref{fig8} and \ref{fig9} a comparison of the first order, the sum of
the first and second order, and the resonant tunneling approximation 
\cite{SS,KSS1} (where the cutoff is adjusted at $E_{\rm C}$) is displayed.
The deviation from sequential tunneling is significant and of the order 
$20\%$. 
The agreement with the resonant tunneling approximation provides a
clear criterium for the choice of the bandwidth cutoff. 
We have checked the significance of third order terms $\sim \alpha_0^3$ by using 
the resonant tunneling formula \cite{SS,KSS1} and exact results for the average 
charge in third-order at zero temperature \cite{Gra}. 
For the parameter sets used in the figures, the deviations to the sum of first 
and second order terms were smaller than about $2\%$. 
Therefore, at not too low temperatures, second-order perturbation theory is a 
good approximation even if the tunneling resistance approaches the quantum 
resistance.

\section{Comparison with renormalization group results}\label{section rg}

Renormalization group techniques have been applied for the single-electron box 
(which is equivalent to the transistor at zero bias voltage) including two
charge states \cite{Mat,Fal-Schoen-Zim}.
Starting with some initial high-energy cutoff (which is of the order of the 
charging energy $E_{\rm C}$), one integrates out the high-energy modes.
This leads to a renormalization of the system parameters.
This procedure has to be performed until the cutoff reaches the largest energy 
scale $\omega_C$ which is important for the lowest order tunneling processes.
This leads to
\begin{equation}
  {\tilde{\alpha}\over \alpha_0} = {\tilde{\Delta}\over \Delta_0}
  = {1\over 1+2\alpha_0 \ln ( E_C/\omega_C)}
\end{equation}
Using the resonant tunneling approximation \cite{SS,KSS1} we generalized this 
to non-equilibrium situations and find (for $E_{\rm C}$ larger than any other 
energy scale)
\begin{eqnarray}
        \tilde{\alpha} &=& { \alpha_0 \over
        1 + 2 \sum_{\rm r} \alpha_0^{\rm r} \left[ \ln \left( {\beta E_{\rm C}
        \over 2\pi} \right) - {\rm Re} \, \Psi \left( 
        i{\beta (\tilde{\Delta} -\mu_{\rm r}) \over 2\pi} \right) \right] }
        \\
        \tilde{\Delta} &=& { \Delta_0 + 2 \sum_{\rm r} \alpha_0^{\rm r} \, 
        \mu_{\rm r} \left[ \ln \left( {\beta E_{\rm C}\over 2\pi} \right) 
        - {\rm Re} \, \Psi \left( i{ \beta (\tilde{\Delta} -\mu_{\rm r}) \over 
        2\pi } \right)\right] 
        \over
        1 + 2 \sum_{\rm r} \alpha_0^{\rm r} \left[ \ln \left( {\beta E_{\rm C}
        \over 2\pi} \right) - {\rm Re} \, \Psi \left( i{ \beta (\tilde{\Delta} -
        \mu_{\rm r})\over 2\pi } \right) \right] } \, .
\end{eqnarray}
But in both cases, the result is rather qualitatively, and the constants in the 
arguments of the logarithms remain unknown.
Here, in the cotunneling theory, we have taken all relevant charge states into 
account, i.e., the cutoff is provided naturally by the energy of the adjacent 
charge states and has not to be introduced by hand.
Therefore, we can extract the complete arguments from Eqs.~(\ref{current2a}) 
and (\ref{current3a}),
\begin{eqnarray}
        \tilde{\alpha} &=& \alpha_0 - 2 \alpha_0 \sum_{\rm r} \alpha_0^{\rm r} 
        \left[ -1+\ln \left( {\beta E_{\rm C}\over \pi} \right) -{\partial \over 
        \partial \Delta_0} \left( (\Delta_0-\mu_{\rm r}) {\rm Re} \, \Psi \left( 
        i{ \beta (\Delta_0 -\mu_{\rm r}) \over 2\pi } \right) \right) \right]
        \\
        \tilde{\Delta} &=& \Delta_0 - 2 \sum_{\rm r} \alpha_0^{\rm r} 
        (\Delta_0 -\mu_{\rm r}) \left[ 1 + \ln \left( {\beta E_{\rm C}\over \pi}
        \right) - {\rm Re} \, \Psi \left( i{ \beta (\Delta_0 -\mu_{\rm r}) \over
        2\pi } \right)\right] \, .
\end{eqnarray}
In the following we discuss different limits.

\subsection{Linear Response}

At zero bias voltage the system is ``on resonance'' if $\beta\Delta_0 \ll 1$
and ``off resonance'' if $\beta\Delta_0 \gg 1$.

While the resonant tunneling formula yields ``on resonance''
\begin{equation}
        {\tilde{\alpha} \over  \alpha_0 } = {\tilde{\Delta} \over  \Delta_0 }
        ={ 1 \over
        1 + 2 \alpha_0 \left[ \gamma + \ln \left( {\beta E_{\rm C}\over 2\pi} 
        \right) \right] } 
\end{equation}
we get from cotunneling
\begin{eqnarray}
        \tilde{\alpha} &=& \alpha_0 \left[ 1 -2\alpha_0 \left( 
        -1+\gamma+\ln \left( {\beta E_{\rm C}\over \pi} \right) \right) \right]
        \\
        \tilde{\Delta} &=& \Delta_0 \left[ 1 -2\alpha_0 \left( 
        1+\gamma+\ln \left( {\beta E_{\rm C}\over \pi} \right) \right) \right] 
        \, .
\end{eqnarray}
We see that the temperature is the relevant energy scale which determines the
renormalization.

In the opposite case, i.e., ``off resonance'' the resonant tunneling 
approximation leads to
\begin{equation}
        {\tilde{\alpha} \over \alpha_0 } = {\tilde{\Delta} \over \Delta_0 }
        ={ 1 \over
        1+2 \alpha_0 \ln \left( {E_{\rm C}\over \tilde{\Delta}} \right) } \, ,
\end{equation}
and the cotunneling yields
\begin{eqnarray}
        \tilde{\alpha} &=& \alpha_0 \left[ 1 -2\alpha_0 \left( 
        -2+\ln \left( {2 E_{\rm C}\over \Delta_0} \right) \right) \right]
        \\
        \tilde{\Delta} &=& \Delta_0 \left[ 1 -2\alpha_0 
        \left( 1 + \ln \left( {2 E_{\rm C}\over \Delta_0} \right) \right) 
        \right] \, .
\end{eqnarray}
Now, the level splitting itself provides the energy at which the renormalization
procedure stops.

The conductance as well as the energy gap is reduced during the renormalization, 
i.e., the system is effectively ``closer'' at the resonance but with higher
tunnel barriers than the bare system.
As a consequence, $G_2^{(2)}$ is negative and $G_3^{(2)}$ is positive.

\subsection{Nonlinear Response}

At finite bias voltage a new energy scale comes into play.
Furthermore, since the Fermi levels of the reservoirs are separated, the relevant
energy scale may depend on which reservoir is responsible for the 
renormalization.
For $\mu_{\rm R} - \mu_{\rm L}=eV \gg k_B T$, being on resonance with the 
left Fermi level means that $k_{\rm B}T \gg |\Delta_0 - \mu_{\rm L}|$, i.e., 
that again the temperature provides the dominant energy scale for the left 
reservoir, but $|\Delta_0-\mu_{\rm R}| \approx eV$ is the relevant energy for 
the right one.

The resonant tunneling formula leads to 
\begin{eqnarray}
        \tilde{\alpha} &=& { \alpha_0 \over 1 + 2 \alpha_0^{\rm L} \, 
         \left[ \gamma + \ln \left( {\beta E_{\rm C}\over 2\pi} \right) \right]
        + 2 \alpha_0^{\rm R} \, \ln \left( {E_{\rm C}\over eV} 
        \right) }
        \\
        \tilde{\Delta} - \mu_{\rm L} &=& { \Delta_0 - \mu_{\rm L} 
        + 2 \alpha_0^{\rm R} \, eV \ln \left( {E_{\rm C}\over eV} \right) 
        \over
        1 + 2 \alpha_0^{\rm L} \, 
         \left[ \gamma + \ln \left( {\beta E_{\rm C}\over 2\pi} \right) \right]
        + 2 \alpha_0^{\rm R} \, \ln \left( {E_{\rm C}\over eV} \right) },
\end{eqnarray}
and the cotunneling gives
\begin{eqnarray}
        \tilde{\alpha} &=& \alpha_0 \left[ 1 - 2\alpha_0^{\rm L}  \left( 
        -1+\gamma+\ln \left( {\beta E_{\rm C}\over \pi} \right) \right)
        - 2\alpha_0^{\rm R}  \left( -2+\ln \left( {2 E_{\rm C} \over eV} 
        \right) \right) \right]
        \\
        \tilde{\Delta} - \mu_{\rm L} &=& \Delta_0 -\mu_{\rm L} 
        -2 \alpha_0^{\rm L} \left( \Delta_0 -\mu_{\rm L} \right) \left[ 1 
        +\gamma +\ln \left( {\beta E_{\rm C}\over \pi} \right) \right]
        +\alpha_0^{\rm R} eV \left[ 1+ \ln \left( {2 E_{\rm C} \over eV}\right)
        \right]
\end{eqnarray}
In the opposite case, i.e., $|\Delta_0 - \mu_{\rm L}| \gg k_{\rm B}T$ and 
$|\Delta_0 - \mu_{\rm R}| \gg k_{\rm B}T$ the resonant tunneling yields
\begin{eqnarray}
        \tilde{\alpha} &=& { \alpha_0 \over
        1 + 2 \sum_{\rm r} \alpha_0^r \, 
        \ln \left( {E_{\rm C}\over |\tilde{\Delta} -\mu_{\rm r}|} \right) } 
        \\
        \tilde{\Delta} &=& { \Delta_0 + 2 \sum_{\rm r} \alpha_0^{\rm r} \, 
        \mu_{\rm r} \ln \left( {E_{\rm C}\over |\tilde{\Delta} -\mu_{\rm r}|} 
        \right) \over 1 + 2 \sum_{\rm r} \alpha_0^{\rm r} \, 
        \ln \left( {E_{\rm C}\over |\tilde{\Delta} -\mu_{\rm r}|} \right) } 
\end{eqnarray}
in comparison to cotunneling,
\begin{eqnarray}
        \tilde{\alpha} &=& \alpha_0 \left[ 1 - 2 \sum_{\rm r} \alpha_0^{\rm r} 
        \left( -2+\ln \left( {2 E_{\rm C}\over |\Delta_0-\mu_{\rm r}|} \right) 
        \right) \right]
        \\
        \tilde{\Delta} &=& \Delta_0 - 2 \sum_{\rm r} \alpha_0^{\rm r} 
        (\Delta_0 -\mu_{\rm r}) \left[ 1 + \ln \left( {2 E_{\rm C}\over 
        |\Delta_0-\mu_{\rm r}|} \right)\right] \, .
\end{eqnarray}

\begin{figure}
\centerline{\psfig{figure=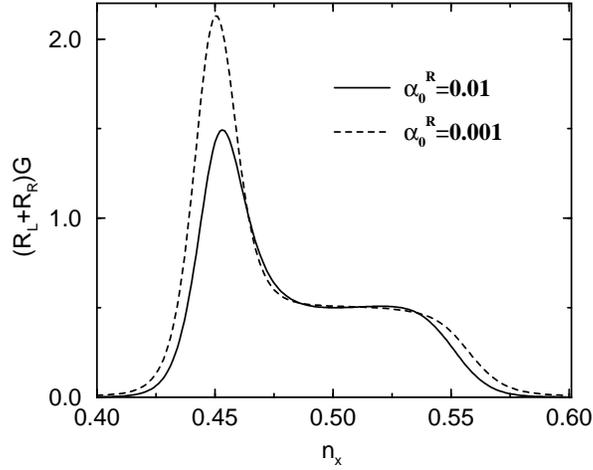,height=7cm}}
\caption{The differential conductance $T/E_{\rm C}=0.01$, $V/E_{\rm C}=0.2$,
        $\alpha_0^{\rm L}=0.0001$, $\alpha_0^{\rm R}=0.01$ (solid line) and 
        $\alpha_0^{\rm R}=0.001$ (dashed line).}
\label{fig10}
\end{figure}

The renormalization of the energy gap differs qualitatively from that in the
linear response regime, since now there is a competition between the 
renormalization of the system towards the left and the right Fermi level.
This effect is displayed in Fig.~\ref{fig10} which shows the differential 
conductance as a function of the external gate charge.
The position at which the system is in resonance with the left Fermi level
is shifted towards the right Fermi level with increasing coupling strength to
the right reservoir.

\section{Linear response for arbitrary temperature}\label{section all temp}

In the last section we discussed the effect of the quantum fluctuations at
low temperature, when Coulomb oscillations are pronounced.
In this case, at most two charge states are occupied classically.
But also at high temperature, the asymptotic behavior shows deviations from the
sequential tunneling result.
Furthermore, for a comparison with experiment, it is necessary to describe also
the crossover between the low and the high temperature regime.
In this section we discuss the linear response, i.e., we set $V=0$, but we allow
for all charge states.

The corrections to the probabilities are then  
\begin{eqnarray}
        p_n^{(1)} = \left( p_n^{(0)} - p_{n-1}^{(0)} \right) \partial \phi_{n-1}
                - \left( p_{n+1}^{(0)} - p_n^{(0)} \right) \partial \phi_n
                - \beta p_n^{(0)} \left( \phi_{n-1} - \phi_n \right)
                \nonumber \\
                + \beta p_n^{(0)} \sum_{n'} p_{n'}^{(0)} 
                \left( \phi_{n'-1} -\phi_{n'}\right)
                - \alpha_0 W \left( 2 p_n^{(0)} - p_{n+1}^{(0)} - p_{n-1}^{(0)}
                  \right)
\end{eqnarray}
which implies
\begin{equation} \label{average charge}
        \langle n \rangle^{(1)}= -{1\over 2 E_{\rm C}}
        {\partial \over \partial n_{\rm x}}
        \left[ \sum_n p_n^{(0)} \left( \phi_{n-1} - \phi_n  \right) \right] \, .
\end{equation}
A systematic perturbative expansion of the partition function (up to order 
$\alpha_0^2$) was performed by Grabert~\cite{Gra}.
The result Eq.~(\ref{average charge}) is identical to his result in order 
$\alpha_0$, which at $T=0$ reads 
$\langle  n \rangle^{(1)}=\alpha_0 \ln [(1+2n_{\rm x})/(1-2n_{\rm x})]$.

Using the definition $S(\Delta_n)=\partial B(\Delta_n) / \partial V$ we find 
the differential conductance 
$G^{(2)}=\partial I^{(2)} / \partial V=\sum_{i=1}^4 G^{(2)}_i$ with
\begin{eqnarray}\label{conductance1}
        G^{(2)}_1&=&
        \int d\omega \, S(\omega) \alpha(\omega)
        \sum_n p_n^{(0)}{\rm Re} \, R_n(\omega)^2 \, ,
        \\ \label{conductance2}
        G^{(2)}_2 &=& - {1\over 2}\sum_n S(\Delta_n)
        \int d\omega \, \alpha(\omega) {\rm Re} \left\{
        \left( p_n^{(0)}+p_{n+1}^{(0)}\right) 
        \left[ R_n(\omega)^2 + R_{n+1}(\omega)^2 \right] 
        \right. \nonumber \\ && \left.
        + \beta p_n^{(0)} R_n(\omega)
        + \beta p_{n+1}^{(0)} R_{n+1}(\omega)
        - \beta \left( p_n^{(0)}+p_{n+1}^{(0)}\right)
        \sum_{n'} p_{n'}^{(0)} R_{n'}(\omega)
        \right\}
        \\ \label{conductance3}
        G^{(2)}_3 &=& - {1\over 2}
        \sum_n  \left( p_n^{(0)}+p_{n+1}^{(0)}\right) 
        {\partial S(\Delta_n) \over \partial \Delta_n} \, 
        \int d\omega \, \alpha(\omega) {\rm Re} \, 
        \left[ R_n(\omega) - R_{n+1}(\omega) \right] 
        \\ \label{conductance4}
        G^{(2)}_4&=& {4\pi^2e^2\over h} \beta \sum_n \int d\omega \,
        p_{n-1}^{(0)} \alpha_{\rm L}^+(\Delta_n+\Delta_{n-1}-\omega) 
        \alpha_{\rm R}^+(\omega) {\rm Re} \,  R_n(\omega)^2
\end{eqnarray}
In comparison to the previous section, a new term, $G^{(2)}_4$, appears, which is
related to new processes where the island charge is increased or decreased by
$2e$.
Again, all integrals Eq.~(\ref{conductance1}) - (\ref{conductance4}) can be
performed analytically.
While the sequential tunneling result gives
\begin{equation}\label{cond seq}
        {G^{(1)} \over G_{\rm as}} = \sum_n \left[ p_n^{(0)}+p_{n+1}^{(0)} 
        \right] {\beta\Delta_n/2 \over \sinh \beta\Delta_n}
\end{equation}
we find for the cotunneling
\begin{eqnarray}\label{analyt cond 1}
        {G^{(2)}_1\over G_{\rm as}} &=& 
        \sum_n p_n^{(0)} \left\{
        \Delta_{n-1} \partial^2 \phi_{n-1} + \Delta_n \partial^2 \phi_n
        + {\Delta_{n-1} \partial \phi_{n-1} - \Delta_n \partial \phi_n 
        \over E_{\rm C}}
        +{\phi_n -\phi_{n-1}\over E_{\rm C}} \right\}
        \\ \label{analyt cond 2}
        {G^{(2)}_2\over G_{\rm as}} &=& 
        \sum_n \left({\beta\Delta_n/2 \over \sinh \beta\Delta_n} \right)
        \left\{  \left( p_n^{(0)}+p_{n+1}^{(0)}\right) 
        \left[ {\phi_{n-1} - \phi_{n+1} \over E_{\rm C}}
        + \partial \left( 2\phi_n + \phi_{n-1} + \phi_{n+1} \right) \right] 
        \right. \nonumber \\ && \left.
        -\beta p_n^{(0)} [ \phi_{n-1}-\phi_n]
        -\beta p_{n+1}^{(0)} [ \phi_n -\phi_{n+1}]
        +\beta \left( p_n^{(0)}+p_{n+1}^{(0)}\right) \sum_{n'} p_{n'}
        [\phi_{n'-1} - \phi_{n'} ] \right\}
        \\ \label{analyt cond 3}
        {G^{(2)}_3\over G_{\rm as}} &=& 
        \sum_n \left[ {\partial \over \partial \Delta_n}
        \left({\beta\Delta_n/2 \over \sinh \beta\Delta_n} \right) \right]
        \left( p_n^{(0)}+p_{n+1}^{(0)}\right) 
        [2\phi_n - \phi_{n-1} - \phi_{n+1} ] 
        \\ \label{analyt cond 4}
        {G^{(2)}_4\over G_{\rm as}} &=& \beta
        \sum_n {  p_{n-1}^{(0)}+p_{n+1}^{(0)} \over \sinh 
        \beta \left( \Delta_{n-1}+\Delta_n \right)}
        \left\{ - \Delta_{n-1} \partial \phi_n - \Delta_n \partial \phi_{n-1}
        +{\Delta_{n-1}+\Delta_n \over 2 E_{\rm C}} [\phi_n - \phi_{n-1}]
        \right\} \, .
\end{eqnarray}

\begin{figure}
\centerline{\psfig{figure=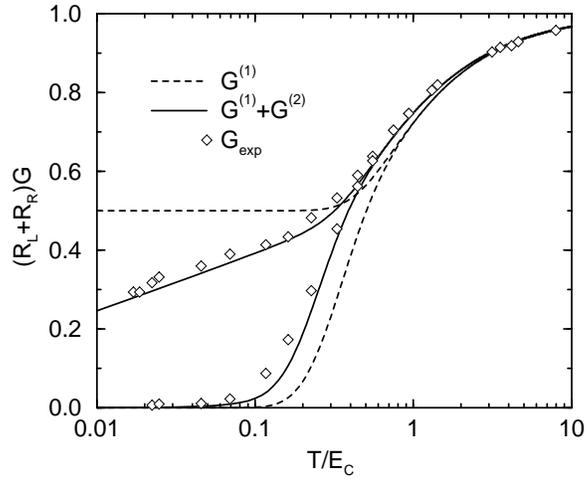,height=7cm}}
\caption[a]{Maximal and minimal linear conductance for $E_{\rm C}=1 K$ and 
        $\alpha_0=0.063$.
        The dashed line is sequential tunneling, the solid line sequential plus
        cotunneling.
        The data points are experimental data from Ref.~\onlinecite{esteve}.}
\label{fig11}
\end{figure}

In Fig.~\ref{fig11} we compare our results with recent experiments \cite{esteve}. 
The temperature dependence of the Coulomb oscillations were measured for a
sample with a conductance $\alpha_0=0.063$.
Our results in second-order
perturbation theory agree perfectly in the whole temperature range. 
For stronger tunneling higher-order effects such as the inelastic resonant 
tunneling \cite{KSS1} would be relevant. 

We emphasize, that only  bare values for $\alpha_0$ and $E_{\rm C}$ have 
been used here, as determined directly in the experiment. 
In contrast, the resonant tunneling approximation with bare values of the 
charging energy would lead to a deviation from the experiment by about $10\%$.
Thus, the inclusion of higher-order charge states within second-order 
perturbation theory, as presented in this paper, is important for comparison
with experiments.

While at resonance the new terms are crucial, the Coulomb blockade 
regime is sufficiently described by Eq.~(\ref{current1}) which yields a good
agreement between theory and experiment.

\section{Asymptotic behaviour}\label{section asymptotic}

In linear response, the whole temperature regime is covered by 
Eqs.~(\ref{cond seq}) - (\ref{analyt cond 4}).
At high and low temperatures these formulas become very simple.

\subsection{High-temperature expansion}

At high temperature ($\beta E_{\rm C} \ll 1$), Coulomb oscillations are 
washed out, i.e., there is no gate charge dependence of the conductance, and 
using Euler-MacLaurin formula we can replace the sum in Eq.~(\ref{cond seq}) by 
an integral.
In this way we average the conductance over all possible gate charges,
loosing the information about the difference between minimal and maximal 
conductance.
In linear response we have $p_n^{(0)}=\exp[-\beta E_{\rm C} (n-n_{\rm x})^2]/Z$. 
We expand $(\beta \Delta_n) / \sinh (\beta \Delta_n)$ in $\beta \Delta$ and 
perform the integral which approximates Eq.~(\ref{cond seq}).
This yields $G^{(1)}/ G_{\rm as}= 1 - \beta E_{\rm C}/ 3 
+ (\beta E_{\rm C})^2 /15 +{\cal O}((\beta E_{\rm C})^3)$.

For the cotunneling contribution we expand the digamma functions which enter
$\phi_n=\alpha_0 \Delta_n {\rm Re} \,\Psi\left(i {\beta \Delta_n \over 2\pi} 
\right)$ using ${\rm Re} \, \Psi (ix)={\rm Re} \, \Psi (1+ix)= -\gamma+ 
\zeta(3) x^2 - \zeta(5) x^4 + {\cal O} (x^6)$, and we find 
$G_1^{(2)} = G_2^{(2)} = G_4^{(2)} = 2\alpha_0(\beta E_{\rm C})^2 \zeta (3)/ 
\pi^2$ and $G_3^{(2)} = 0$, i.e.,
\begin{equation}\label{high temperature}
        {G\over G_{\rm as}}= 1 -{\beta E_{\rm C} \over 3} + (\beta E_{\rm C})^2
        \left[ {1\over 15}+ {6 \zeta (3) \over \pi^2} \alpha_0  \right]
        + {\cal O}\left( (\beta E_{\rm C})^3 \right) \, .
\end{equation}  
This result was also derived in Refs.~\onlinecite{Gol-Zai/GKSSZ,Goe-Gra}.

\subsection{Low-temperature expansion}

At low temperature ($\beta E_{\rm C} \gg 1$), Coulomb oscillations appear.
The maximal conductance is reached at half integer values of $n_{\rm x}$.
In sequential tunneling the maximal conductance saturates at one half of the 
asymptotic high temperature value. 
The quantum fluctuations, however, lead to a further reduction.
Approximating the digamma function by ${\rm Re} \, \Psi (ix)= \ln |x|$ for 
$x\neq 0$, we find 
$G_{1,\rm max}^{(2)}/ G_{\rm as}= -\alpha_0,\,\,\,
G_{2,\rm max}^{(2)}/G_{\rm as} = \alpha_0 \left[ 1-\gamma -\ln \left( 
{\beta E_{\rm C} \over \pi} \right) \right]$, and 
$G_{3,\rm max}^{(2)}=G_{4,\rm max}^{(2)}=0$, i.e., in total
\begin{equation}\label{low temperature}
        {G_{\rm max} \over G_{\rm as}}= {1\over 2}-\alpha_0 \left[ \gamma 
        +\ln \left( {\beta E_{\rm C} \over \pi} \right)\right]
        +{\cal O}\left( \alpha_0^2 \right)
\end{equation}
with $\gamma$ being Euler's constant and $G_{\rm as}=1/(R_L+R_R)$ the 
asymptotic high temperature limit.
The peak conductance depends logarithmically on temperature. 
This result may be interpreted as a renormalization of $G_{\rm as}$ or 
$\alpha_0$ \cite{SS,KSS1,KSS3,Mat,Fal-Schoen-Zim}.
It shows a typical logarithmic temperature dependence since, at least
in the equilibrium situation, the low-energy behavior of the system
is expected to be that of the multichannel Kondo model \cite{Mat}.

\begin{figure}
\centerline{\psfig{figure=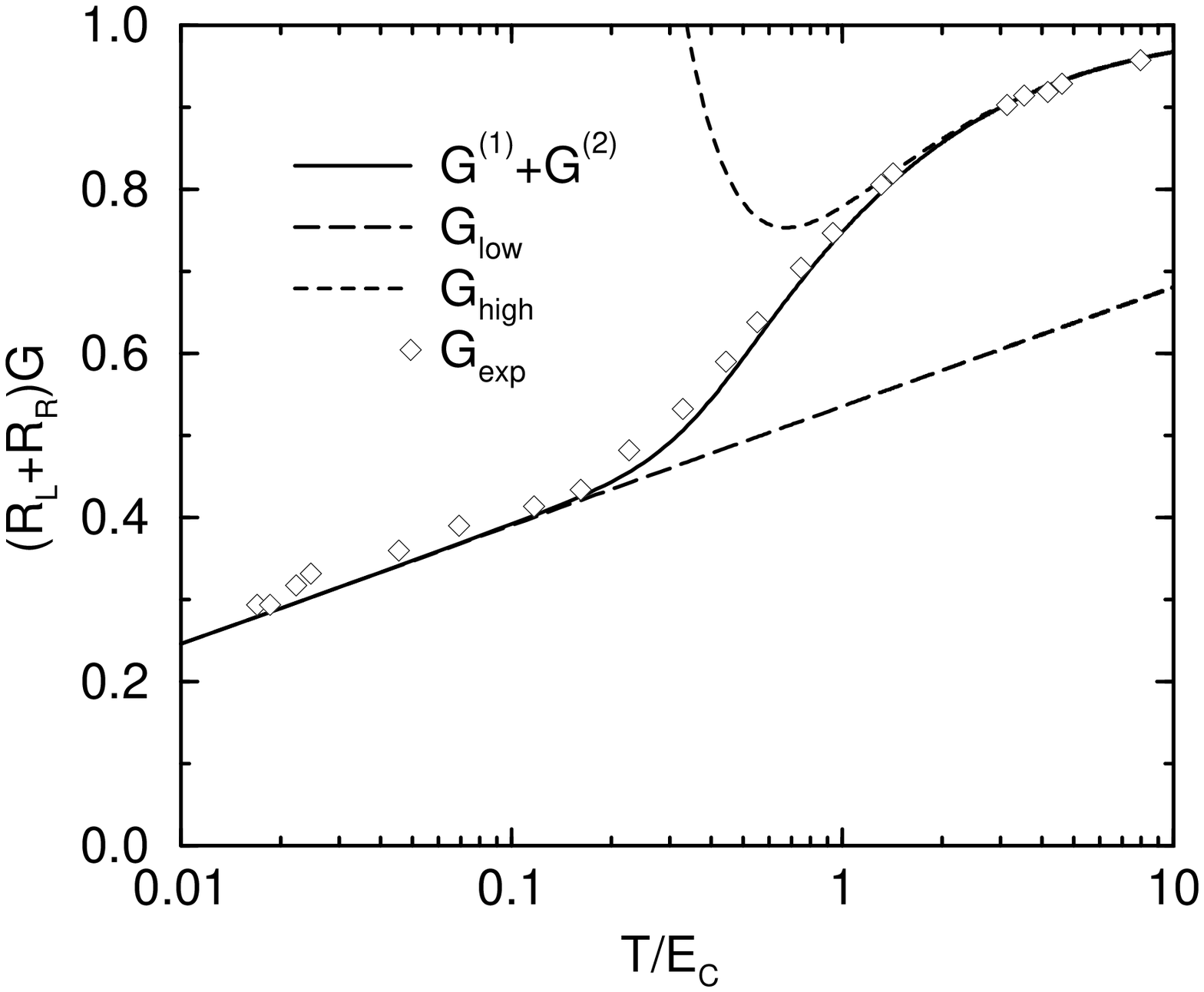,height=7cm}}
\caption[a]{Maximal linear conductance for $E_{\rm C}=1 K$ and $\alpha_0=0.063$.
        The solid line (full first plus second order result) interpolates 
        between the long-dashed line (low-temperature expansion) and the dashed
        line (high-temperature expansion).
        The data points are experimental data from Ref.~\onlinecite{esteve}.}
\label{fig12}
\end{figure}

In Fig.~\ref{fig12} we compare the limiting expressions 
Eq.~(\ref{high temperature}) and (\ref{low temperature}) with the full 
second-order result Eq.~(\ref{cond seq}) - (\ref{analyt cond 4}).
Our theory interpolates perfectly between the logarithmic behavior
at low temperature and the high-temperature result.

\section{Conclusion}

In conclusion we have evaluated the current of the single-electron transistor 
consistently up to second-order perturbation theory (cotunneling) at arbitrary 
temperature and transport voltage.
The approach is free of divergences and provides cutoff-independent results. 
Out of resonance we recover the usual ''cotunneling'' contribution at low 
temperature \cite{Ave-Naz}.
At resonance, however, we find new terms which are significant for 
experimentally realistic parameters.
They describe the renormalization of the dimensionless conductance and
of the energy gap of two adjacent charge states.
Both parameters are reduced in comparison to the unrenormalized value and depend
logarithmically on temperature and bias voltage.
We can extract the arguments of the logarithms which are not provided by 
renormalization group techniques and which are important for comparison with 
experiments.
Furthermore, we include processes in which two electrons enter or leave the 
island coherently.
The corresponding term is important to describe the deviation of the sequential
tunneling result at high temperatures.

We have derived analytical expressions at low temperatures, including 
nonequilibrium effects.
In addition we have studied the linear response regime for arbitrary 
temperature.
We are, thus, able to describe the crossover between the Coulomb blockade
and classical high-temperature regime. 
A comparison with experiments shows good quantitative agreement.

We like to thank D. Esteve, H. Grabert, and P. Joyez for stimulating and
useful discussions. 
Our work was supported by the ``Deutsche Forschungsgemeinschaft'' as part of 
``SFB 195''.

\begin{references}
\bibitem{Ave-Lik}
D.V. Averin and K.K. Likharev, in {\it Mesoscopic Phenomena in Solids},
ed. B.L. Altshuler et al. (Elsevier, 1991).

\bibitem{Gra-Dev}
{\it Single Charge Tunneling}, NATO ASI Series {\bf 294},  
H. Grabert and M.H. Devoret, eds., (Plenum Press, 1992).

\bibitem{Sch-Uebersicht}
 G. Sch\"on, {\sl Single-Electron Tunneling}, Chapter 3 in {\sl Quantum 
Transport and Dissipation}, T. Dittrich, P. H\"anggi, G.-L. Ingold, B. Kramer, 
G. Sch\"on, and W. Zwerger, (Wiley-VCH Verlag, 1998). 

\bibitem{Ave-Naz}
D.V. Averin and A.A. Odintsov, Phys. Lett. A {\bf 140}, 251 (1989);
D.V. Averin and Yu.V. Nazarov, Phys. Rev. Lett. {\bf 65}, 2446 (1990);
{\it ibid.} in Chapter 6 in Ref.~[2].

\bibitem{Exp-Cot}
L.J. Geerligs, D.V. Averin, and J.E. Mooij, Phys. Rev. Lett. {\bf 65},
3037 (1990); U. Meirav et al.,
Phys. Rev. Lett. {\bf 65}, 771 (1990); Chapter 3,6 in Ref.~[2].

\bibitem{SS}
H. Schoeller and G. Sch\"on, Phys. Rev. B {\bf 50}, 18436 (1994); 
Physica B {\bf 203}, 423 (1994). 

\bibitem{KSS1}
J. K\"onig, H. Schoeller, and G. Sch\"on, Europhys. Lett. {\bf 31}, 31 (1995); 
and in {\it Quantum Dynamics of Submicron Structures}, eds. H. A. Cerdeira 
{\it et al.}, NATO ASI Series E, Vol. 291 (Kluwer, 1995), p.221.

\bibitem{KSS3}
J. K\"onig, H. Schoeller, and G. Sch\"on, Phys. Rev. Lett. {\bf 78}, 4482 (1997).

\bibitem{Laf}
P. Lafarge et al., Z. Phys. B  {\bf 85}, 327 (1991).

\bibitem{Mat}
K.A. Matveev, Sov. Phys. JETP {\bf 72}, 892 (1991),
[Zh. Eksp. Teor. Fiz. {\bf 99}, 1598 (1991)].

\bibitem{Gol}
D.S. Golubev and A.D. Zaikin, Phys. Rev. B {\bf 50}, 8736 (1994).

\bibitem{Fal-Schoen-Zim}
G. Falci, G. Sch\"on, and G.T. Zimanyi, Phys. Rev. Lett. {\bf 74}, 3257 (1995).

\bibitem{Gra}
H. Grabert, Phys. Rev. B  {\bf 50}, 17364 (1994).

\bibitem{KSS2}
J. K\"onig, H. Schoeller, and G. Sch\"on, Phys. Rev. Lett. {\bf 76}, 1715 (1996).

\bibitem{KSSS} 
J. K\"onig, J. Schmid, H. Schoeller, and G. Sch\"on, Phys. Rev. B {\bf 54}, 
16820 (1996).

\bibitem{esteve}
P. Joyez, V. Bouchiat, D. Esteve, C. Urbina, and M.H. Devoret,
Phys. Rev. Lett. {\bf 79}, 1349 (1997).

\bibitem{com}
The master equation is equivalent to $I(n)=I(n+1)=const.$.
To determine the constant we consider the limit $n\rightarrow \infty$ and find
$I(n)=0$.

\bibitem{com3}
Usually, the value of the integral is approximated by replacing the denominators 
with an $\omega$-independent term or by adding a constant finite life-time in 
the denominator, see e.g. Y.V. Nazarov, J. Low Temp. Phys. {\bf 90}, 77 (1993); 
D.V. Averin, Physica {\bf 194-196}, 979 (1994);
P. Lafarge and D. Esteve, Phys. Rev. B {\bf 48}, 14309 (1993). 

\bibitem{Gol-Zai/GKSSZ}
D.S. Golubev and A.D. Zaikin, JETP Lett. {\bf 63}, 1007 (1996), [Zh. Eksp. Teor. 
Fiz. Pis'ma Red. {\bf 63}, 953 (1996)];\\
D.S. Golubev, J. K\"onig, H. Schoeller, G. Sch\"on, and A.D. Zaikin, 
Phys. Rev. B {\bf 56}, 15782 (1997).

\bibitem{Goe-Gra}
G. G\"oppert and H. Grabert, unpublished.

\end{references}
\end{document}